\documentclass{article}
\usepackage{graphicx} % Required for inserting images
\usepackage{enumitem}
\usepackage[utf8]{inputenc}
\usepackage{booktabs}
\usepackage[T1]{fontenc}
\usepackage{graphicx}
\usepackage{authblk}
\usepackage{amsmath,amsfonts,amssymb}
\usepackage{mathtools}
\usepackage{hyperref}
\usepackage{soul,color}
\usepackage{geometry}
\usepackage{tabularx} 
\usepackage{pdflscape}
\usepackage{natbib}
\usepackage{longtable}
\usepackage{booktabs}
\usepackage[table,xcdraw]{xcolor}
\usepackage{url}
\usepackage{comment}
\usepackage{tabularx}

\newcounter{siSubsection}
\renewcommand{\thesubsection}{SI.\arabic{siSubsection}}

\newcommand{\sisubsection}[1]{
  \refstepcounter{siSubsection}
  \subsection*{#1}
  \addcontentsline{toc}{subsection}{\thesubsection \quad #1}
}

\newcommand{\startupexposure}{AISE }

\title{Follow the money: a startup-based measure of AI exposure across occupations, industries and regions}
\author{Enrico Maria Fenoaltea, Dario Mazzilli, Aurelio Patelli, Angelica Sbardella, Andrea Tacchella, Andrea Zaccaria, Marco Trombetti, Luciano Pietronero}
\begin{document}

\maketitle

\begin{abstract}

The integration of artificial intelligence (AI) into the workplace is advancing rapidly, necessitating robust metrics to evaluate its tangible impact on the labour market. Existing measures of AI occupational exposure largely focus on AI's theoretical potential to substitute or complement human labour on the basis of technical feasibility, providing limited insight into actual adoption and offering inadequate guidance for policymakers. 

To address this gap, we introduce the AI Startup Exposure (AISE) index—a novel metric based on occupational descriptions from O*NET and AI applications developed by startups funded by the Y Combinator accelerator. Our findings indicate that while high-skilled professions are theoretically highly exposed according to conventional metrics, they are heterogeneously targeted by startups. Roles involving routine organizational tasks—such as data analysis and office management—display significant exposure, while occupations involving tasks that are less amenable to AI automation due to ethical or high-stakes, more than feasibility, considerations — such as judges or surgeons— present lower AISE scores. Moreover, geographically, AI exposure is concentrated in knowledge-intensive metropolitan areas like San Francisco and Seattle, while service-oriented sectors exhibit greater exposure compared to agriculture and construction. 
By focusing on venture-backed AI applications, our approach offers a nuanced perspective on how AI is reshaping the labour market. It challenges the conventional assumption that high-skilled jobs uniformly face high AI risks, highlighting instead the role of today's AI players' societal desirability-driven and market-oriented choices as critical determinants of AI exposure. Contrary to fears of widespread job displacement, our findings suggest that AI adoption will be gradual and shaped by social factors as much as by the technical feasibility of AI applications. This framework provides a dynamic, forward-looking tool for policymakers and stakeholders to monitor AI's evolving impact and navigate the changing labour landscape.
\end{abstract}

\section*{Introduction}

 The public and scholarly debate on the employment effects of the new wave of AI developments, especially since the deployment of generative AI, is highly polarized and presents contrasting views on their potential risks and benefits. While some fear a jobless future \citep{marken2023three}, others foresee job creation and complementarity between human and automated tasks, potentially leading to a productivity boost due to AI \citep{georgieff2021artificial}. Beyond the differing views, there is consensus among academics and policymakers that the current wave of artificial intelligence fundamentally differs from previous technological shifts \citep{staneva2023measuring}, distinguished by its unprecedented ability to mimic human reasoning and creativity across a wide range of applications. This unique capability to perform complex, non-routine tasks -- exemplified by large language models like OpenAI’s ChatGPT and Anthropic’s Claude -- coupled with its rapid pace of adoption and improvement \citep{jung2024transformed} positions AI as potentially one of the “most significant general-purpose technology of our era” \citep{brynjolfsson2017artificial}, with far-reaching implications for the economy at large, and heterogeneous effects across jobs, sectors, and countries \citep{acemoglu2021harms,georgieff2021artificial}.

As AI applications continue to emerge, extensive research has already focused on its impact on the labour market. There is growing evidence that AI is reshaping labour demand \citep{acemoglu2020robots,acemoglu2022artificial,albanesi2023new,fleming2019robots}, yet the overall employment effects remain unclear due to the large uncertainty on how these rapidly evolving technical developments will be adopted and deployed \citep{acemoglu2022artificial,acemoglu2021harms,autor2022labor}. Empirical findings vary widely, from low disruption \citep{brynjolfsson2017can,brynjolfsson2018can,gmyrek2023generative} to high displacement potential \citep{eloundou2024gpts}. Some studies show AI having complementary effects \citep{gmyrek2023generative,ellingrud2023generative} and driving productivity gains, particularly in high-skilled occupations \citep{brynjolfsson2023turing,Noy2023,Peng2023}, while others remain inconclusive about whether AI leads to complementarity or substitution \citep{felten2019occupational,felten2021occupational}. Concerns have emerged about AI adoption outpacing the labour market's ability to adapt \citep{brynjolfsson2018can}, the quality of new jobs created \citep{autor2022labor,green2023artificial}, and AI uneven impact across sectors and regions \citep{frank2019toward}. \cite{acemoglu2023can} and \cite{autor2022new} argue that AI’s emphasis on automation, rather than augmenting human tasks, risks further stagnating productivity, wages, and labour demand, while deepening income inequality. To mitigate these risks, they advocate for a human complementarity approach supported by an appropriate set of government policies  \citep{autor2022new, autor2024applying,acemoglu2023can}. Additionally, the diminished worker voice due to AI powered monitoring and surveillance, the dominance of private actors in the AI race, and the absence of clear legislative frameworks have sparked broader concerns about AI societal impact \citep{acemoglu2021harms, autor2022labor, staccioli2024worker}.

While the debate on AI labour market and societal impact lacks consensus, the literature broadly agrees that, unlike previous ICT-based technological changes, AI exposure is highest among high-skilled, white-collar workers \citep{eloundou2024gpts,felten2019occupational,webb2019impact}. AI in fact primarily targets clerical work but affects both routine and non-routine cognitive tasks, whereas manual, operational, and technical tasks are comparatively less exposed.

One of the primary methodological and conceptual challenges for understanding the impact of AI on labour markets is the empirical identification of occupational AI exposure. %By building on the task framework pioneered by \cite{autor2003skill}, in recent years, several approaches aimed at estimating the average AI exposure of an occupation’s content in terms of the overlap between AI capabilities and occupational tasks \citep{brynjolfsson2018can,eloundou2024gpts,frey2017future}, or abilities \citep{felten2019occupational,felten2021occupational,felten2023will,martinez2020does,tolan2021measuring} have been proposed \citep{staneva2023measuring}.
Building on the task framework pioneered by \cite{autor2003skill}, several recent approaches estimate the average AI exposure of occupations based on the overlap between AI capabilities and occupational tasks or abilities \citep{brynjolfsson2018can,eloundou2024gpts,frey2017future, felten2019occupational,felten2021occupational,felten2023will,martinez2020does,tolan2021measuring}.
Among these, the AI Occupational Exposure (AIOE) index proposed by Felten et al. \citeyear{felten2019occupational,felten2021occupational} is becoming a standard in the literature and serves as a benchmark for our analysis -- see the Methods section for more details on its construction. More recently, AI-assisted approaches have emerged, using Large Language Models (LLMs) to assess occupational exposure, as demonstrated by \cite{gmyrek2023generative}, who estimate task-level scores of occupational exposure to AI using ChatGPT-4, and \cite{eloundou2024gpts}, who combine expert opinions with ChatGPT-4 classifications to quantify the impact of Generative Pre-trained Transformers (GPTs) on the US labour market.
 Finally, \cite{webb2020,meindl2021exposure, sousa2023artificial} have proposed patent-based methods, employing NLP techniques to analyze AI patent texts' similarity directly to occupational descriptions. 

In these works, AI exposure is generally determined by three main steps: (i) selection of a set of relevant AI applications -- such as language modeling or image recognition -- either arbitrarily or based on AI benchmarks\footnote{AI benchmarks are standardized tests designed to evaluate AI performance in relevant domains such as visual reasoning and reading comprehension.}; (ii) assessment of the potential for task or ability substitution in various occupations using information via expert judgment, crowd-sourcing platforms, or natural language processing (NLP) techniques, and leveraging detailed descriptions from the O*NET occupational database; (iii) definition of the occupational AI exposure index as the share of the occupation’s bundle of tasks/abilities that the AI applications are capable to substitute for.

 All these efforts, while extremely valuable, share some shortcomings. First, relying on expert or crowd-sourced evaluations of AI capabilities may lead to non-reproducible, subjective estimates. This is partially mitigated by the approaches relying on AI patents, which are more quantitative and less subjective. However, a significant disadvantage is that patents may not cover most AI applications since, as many other software advancements, they are not often patented, and patents may not allow to map in most recent advancements as usually there is a lag between the filing date and the time at which they are observed in patent databases \citep{rudyk2015climate}.
Second, all these indices, irrespective of how they are built, inherently measure the \textit{potential} AI exposure, not \textit{actual} adoption within firms and industries \citep{guarascio2023artificial,svanberg2024beyond}. 
With the exception of \cite{svanberg2024beyond}, who take into account the \textit{economic attractiveness} of automating computer vision, none of these approaches directly include information on the technical feasibility of AI implementation, the economic viability, and the social desirability of adopting AI systems. This may limit their predictive accuracy and usefulness for guiding policy planning. In fact, as the actual diffusion of AI is still in its early stages, no strong evidence of labour substitution seems to emerge for occupations considered most exposed to AI according to existing indices  \citep{albanesi2023new,barbieri2020testing,green2023artificial,mondolo2022composite}. 

 To overcome the \textit{potential} nature of AI exposure metrics and building on the existing approaches in the literature, especially on those relying on LLMs such as \cite{eloundou2024gpts} and \citep{gmyrek2023generative}, in the present paper we propose a novel occupational AI exposure index, the \textit{Occupational AI Startup Exposure} (\startupexposure). AISE is aimed at measuring the near-future, actual exposure of occupations by proxying AI innovations with AI applications developed by the startups funded by Y Combinator Management, a US-based venture capital firm and startup accelerator.\footnote{Y Combinator (YC), founded in 2005, is a leading startup accelerator that has funded over 4,000 startups, with a total combined valuation exceeding \$600 billion. It typically invests \$500,000 in exchange for 7\% equity, running two batches each year that last for three months. YC is renowned for its vast alumni network and mentorship program has produced more than 50 unicorn companies, including Airbnb and Dropbox.} 
 In practice, AISE assigns an exposure score to each occupation by leveraging Meta’s Llama3 state-of-the-art open-weight large language model to asses the similarity between O*NET job descriptions and the descriptions of the AI applications developed by Y Combinator-funded startups -- available from Y Combinator website. 

According to AISE, the occupations displaying highest exposure are General office clerks, Data scientists, Computer and information systems managers, and Market research analysts and marketing specialists. These roles typically involve programming, information processing, or organizational tasks that are increasingly targeted by AI startups. In contrast, Athletes and sports competitors, Magistrate judges, and Pediatric surgeons present low AISE scores. These occupations have more diversified skill sets and often involve tasks that are less amenable to AI automation due to physical, ethical, or high-stakes considerations.
Overall, and in agreement with the “reverse skill-bias” predicted by Acemoglu \cite{acemoglu2020unpacking}, our findings indicate that high-skilled, high-education jobs display the highest AI exposure, albeit with some interesting detours from standard exposure indices. 

When comparing our exposure rankings with that of \cite{felten2021occupational}’s AIOE, we observe that jobs with both low AIOE and Occupational AISE tend to be composed primarily of manual tasks. As AIOE increases, we detect a heterogeneous pattern of AISE, with several occupations displaying lower exposure from our startup-based index compared to the ability-based AIOE. Interestingly, occupations typically requiring a master's degree or higher, significant experience, and skills with high importance scores are concentrated in the area of low AISE but high AIOE. This heterogeneity can help disentangle jobs with similar abilities but different levels of actual AI exposure. It suggests that despite the theoretical exposure to AI, many high-skill, high-education roles are not currently impacted by AI according to our startup-based measure. This highlights that the necessity for advanced skills and the high-stakes associated with errors in these roles make AI integration less straightforward, even when there is theoretical potential for AI involvement.

 While we remain silent on the net labour market effects of AI and on whether AI acts as a complement or a substitute to human labour, we propose a geographical and a sectoral projection of \startupexposure. Geographically, we observe that knowledge-intensive US metropolitan areas with expanding digital economies and tech industries -- such as San Jose, San Francisco, Austin, and Seattle -- show higher average AISE. In contrast, the Midwest shows lower exposure, likely due to its reliance on manufacturing and agriculture. 
At the sector-level, we find that service-oriented industries requiring higher levels of information processing and education tend to have greater AI exposure. In contrast, sectors like Educational Services and Health Care, which also demand high education levels but involve many high-stakes jobs, exhibit lower exposure. Unsurprisingly, sectors such as construction and agriculture are less frequently targeted by AI startups.

Finally, in the supplementary information (SI), we present a preliminary result demonstrating how the same methodology used for the AISE can be applied to create a Robotic Startup Exposure (RSE) index. This index links descriptions of the products and services developed by robotics-focused AI startups to job descriptions. Our analysis shows that many occupations with low AISE scores have high RSE scores, particularly lower-skill jobs that involve more physical abilities. Interestingly, we also find that occupations with high AISE scores tend to also have high RSE scores. These findings suggest that many robotics startups are integrating AI into their products, and the joint action of AI and robotics may open the way for much more pervasive job disruption, not only in manufacturing occupations but across all occupations. In this way, our approach illustrate how one can study the impact of the integration of AI and robotics, which, unlike the impact of robotics and automation alone \citep{acemoglu2018race, autor2018automation, anton2022labour}, remains under-explored in the literature.

The novel measures of AI and AI-powered robotic exposure we propose are grounded in actual investments in AI and provide a more realistic assessment of job exposure. Our findings in fact mitigate the high-skills catastrophe envisioned by other approaches, even though they suggest that the potential impact of AI robotics may be highly pervasive. Indeed, unlike the abstract nature of AI capabilities found in patent or benchmark datasets, AI startups are funded by venture capital because they propose tangible solutions related to the performance of specific tasks, prioritizing economic viability over potential technological feasibility and capturing societal interest, trust, and the willingness to integrate AI into occupations \citep{glikson2020human}. 
 Moreover, a key advantage of our methodology in constructing the \startupexposure  index is its full reproducibility, as it leverages an open-weight LLM that can be freely and locally executed. \startupexposure   can be easily updated as new AI startups are financed by Y Combinator or other venture capital firms, enabling near real-time tracking of AI investments to inform effective policy development.

 In future works, we plan to expand \startupexposure  and its variants to provide more comprehensive insights into AI’s impact on the labour market. This includes evaluating the effects of AI on employment patterns, wage dynamics, and income distribution across different sectors and regions, particularly in the US and Europe. We also aim to explore how AI adoption shapes occupational structures over time, contributing to a deeper understanding of its broader economic and societal implications.

\section*{Results}
\label{sec:results}
\subsection*{From AI startups to AI exposure}
To quantify the AI Startup Exposure (\startupexposure) of occupations, geographical areas and industries we rely on two textual data sources, O*NET and the description of the startup funded by the accelerator Y Combinator, and link new AI developments with occupational characteristics using Meta state-of-the-art LLM Llama 3.

In practice, we feed Llama 3 with the textual descriptions of (i) over 1000 SOC occupations provided by O*NET\footnote{In the O*NET system, a SOC occupation refers to a job classification that is based on the Standard Occupational Classification (SOC) system. The SOC system is a structure used by the U.S. government to categorize all possible occupations in the L’abiura market} and (ii) almost 1000 AI-tagged startups funded by Y Combinator (see the Methods section), %whose core business involves AI applications, to build a new dataset containing a Startup-based Indicator of Revealed Exposure on AI (\startupexposure)occupational exposure index using an LLM-based methodology. 
and exploit the LLM's linguistic abstraction capabilities to determine whether, for each startup-occupation pair, the startup's AI application can substitute one or more of the essential tasks mentioned in the O*NET short occupational descriptions. %necessary to perform the job as provided in O*NET short occupational descriptions.
We refer to the Methods section for the definition of essential tasks and more details on the empirical strategy implementation. 

%The higher the number of startup AI applications identified as substitutes for one or more essential job tasks, the more we consider that job exposed to AI. 
We thus define our Occupational AI Startup Exposure (Occupational AISE) for each job as the normalized number of startups developing AI applications identified by the LLM as substitutes for one or more of the essential tasks present in the job's O*NET short description. 

Our exposure is based on the interest of the startup ecosystem in developing AI applications aimed at partially or fully automating a job, it therefore indirectly captures the AI application feasibility, cost, and attractiveness assessments by both the single startup and Y Combinator \citep{svanberg2024beyond}.%[cit svanberg et al. somehow].
Therefore, AISE remains neutral regarding whether AI will complement or substitute human L’abiura, it instead reflects the potential transformation of work driven by AI considering the automation of critical job tasks that, if replaced, would significantly modify the job. In fact, by focusing on essential tasks, as described in O*NET's short descriptions, the index highlights how automating these tasks could alter the nature of the job and the skill-set required to perform it.

%In this context, our exposure index measures the interest of the startup market in influencing significantly a particular job, but should not be interpreted as a substitution index. Indeed, for each job, we leverage Llama 3's capabilities to count the number of startups that could replace at least one of its essential tasks, not necessarily all tasks. 
%Forse qui va aggiunto qualcosa enfatizzando ancora di più che non è neanche un indice di esposizione (agnostoco rispetto a complementarità o sostituzione), ma può essere interpretato come un'indice di "trasformazione" del lavoro a causa dell'AI. In fatti stiamo considerando la sostituzione dei task essenziali (presenti nella short descriprion di O*NET), ovvero tasks che, se sostituiti, modificherebbero il lavore e le corrisopondenti skill richeiste
\subsection*{AI Occupational exposure }
To provide an initial impression of where the AI startup market is headed and which insights Occupational AISE can provide, it is interesting to look in detail at which jobs are most and least exposed to AI (an extended list of the most and least exposed occupations according to our index is presented in the SI Tables \ref{tab:1} and \ref{tab:2}). 
The job with the highest Occupational AISE is \textit{General office clerks}, described by O*NET as requiring \textit{Knowledge of office systems and procedures} and which tasks include a \textit{Maintain and update filing, inventory, mailing, and database systems, either manually or using a computer}.
and \textit{Compile, copy, sort, and file records of office activities, business transactions, and other activities}. In view of the rapid advancement of generative AI, which increasingly simplifies and automates the generation and processing of any format of information (text, audio, video), it is not difficult to imagine that a large portion of such tasks mentioned is highly substitutable by startups developing LLMs or multi AI agents systems. Among the AI applications proposed by startups funded by Y Combinator and classified by Llama 3 as potential substitutes for at least one of the essential tasks of a general office clerk, we mention two illustrative cases. The startup \textit{Nowadays} develops an AI-powered event planning copilot capable of organizing large-scale corporate events -- e.g., it can contact venues, negotiate, and handle all administrative tasks; the startup \textit{Quickchat AI} develops a platform to build multilingual AI assistants powered by generative AI models such as GPT that can perform conversational tasks -- e.g., answering phone calls or processing information for organizational tasks. 

Other jobs with high Occupational AISE include \textit{Data scientists}, \textit{Computer and information systems managers}and \textit{Market research analysts and marketing specialists}. According to O*NET, all these jobs require tasks related to programming and information processing or organizational and planning tasks which by the same token are increasingly targeted by AI-based startups.

The analysis of the jobs least exposed to AI, as determined by Occupational AISE, is less straightforward because these occupations present more diversified skill-sets and educational/training requirements. To illustrate, let us consider three jobs with low Occupational AISE scores: \textit{Athletes and sports competitors}, \textit{Magistrate judges}, and \textit{Pediatric surgeons}. 
Athletes, while benefiting from AI tools for performance monitoring and injury prevention, primarily engage in tasks that rely on human physical abilities and their societal value lies intrinsically in their ability to push human limits while remaining human. %For instance, even if a robot could run faster than Usain Bolt, it would not hold the same societal interest.

Judges, in contrast, perform tasks requiring advanced cognitive skills, such as information processing and decision-making in complex contexts -- tasks that in principle generative AI could already perform. However, the high-stakes and ethically charged nature of judicial work poses significant barriers to AI adoption in this field. This challenge is compounded by the ongoing heated debate about the role of AI in legal and judicial contexts \citep{remus2017can, reiling2020courts} and the ethical implications and introduction of undesired biases in the increasing use of AI in the US criminal justice system \cite{surden2019artificial,yamane2020artificial,gordon2021ai}, highlighting the critical role of the "human factor" and personal accountability in judges' rulings, all factors contributing to a limitation to automation, beyond a mere assessment of technological feasibility. %Moreover, even if AI could outperform humans with less bias, society might impose unalterable requirements, because this is a highly regulated field. Even clearly superior AI judges would not be used until explicitly considered by the body of laws.

Finally, pediatricians, like other medical professionals with low Occupational AISE scores, require a combination of manual skills (such as handling instruments or treating patients), cognitive abilities, and social skills. Although some essential tasks, like symptom-based diagnosis, could be and in some contexts are already automated with AI 
 \cite{malik2019overview,rajpurkar2022ai}, the medical field's sensitivity to errors -- due to their potentially dramatic consequences and the accompanying legal and ethical implications -- discourages AI startups from targeting the substitution of critical medical tasks.
 
 These examples demonstrate how our measure of Occupational \startupexposure effectively captures not only the overlap between the human and AI capabilities in performing specific tasks but also the societal attractiveness of such exposure \citep{glikson2020human}. This insight is made possible by the use of data based on concrete attempts and investments, reflecting both the potential and the limitations of AI integration across various professions: AI exposure is not solely driven by technical feasibility, multiple other societal constraints can accelerate, slow down, or even halt AI relevance.

As mentioned n the introduction, we applied a similar methodology to develop an index for Robotic Startup Exposure (RSE), concentrating on the products and services created by AI-robotics startups. The RSE index indicates that the combined impact of AI and robotics is likely to affect a wide range of occupations, including many roles requiring manual and physical skills. However, these findings are preliminary. For more details, refer to the Supplementary Information and Methods sections. In the following discussion, we focus solely on AISE.
 
\subsection*{AISE and AIOE}
\begin{figure}[t!]
	\centering
	\includegraphics[width=1\textwidth]{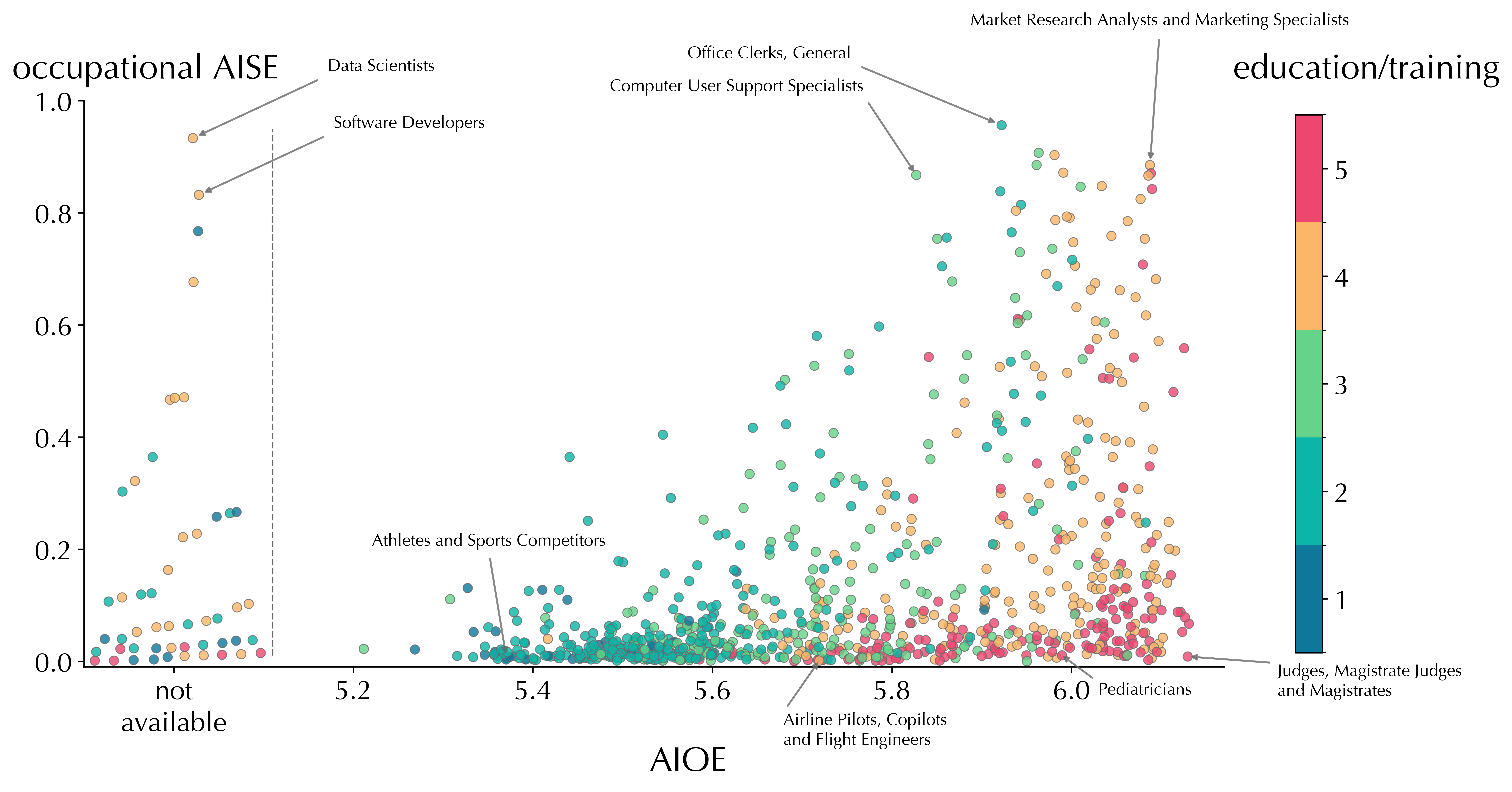}
	\caption{\textbf{Scatter plot of \startupexposure and AIOE values for different jobs of O*NET}. Each point represents a job classified by O*NET and its color represents the corresponding education and training level by the \textit{job zone} feature available in O*NET, as explained in the main text. The left column of points \textit{not available} collects the job with a \startupexposure value but no AIOE.}
	\label{fig:main}
\end{figure}

%To investigate the AI feasibility, 
To externally validate our measure of AI exposure and assess the differences of our startup-based view of the role of new AI developments in the world of work, we compare the Occupational AISE with one of the most used exposure metric in the literature, the AI Occupational Exposure (AIOE) index introduced by \cite{felten2021occupational}. AIOE is built by first quantifying the AI exposure of single O*NET abilities -- e.g., \textit{Deductive reasoning} or \textit{Negotiation} -- through a crowd-sourced survey. Second, by considering an occupation as the bundle of abilities it uses, an occupation AI exposure is measured as the average exposure of the required abilities -- for more details on AIOE and its construction see Section \ref{sec:data&methods}. Therefore, since abilities are personal attributes that can be associated to different occupations, relying on this intermediate layer, AIOE does not capture a job actual AI exposure but its theoretical or potential exposure based on current technological feasibility \citep{tolan2020,staneva2023measuring}.

Figure~\ref{fig:main} shows the \startupexposure vs AIOE relationship for each SOC occupation, color-coded according to O*NET five Job Zones. A Job Zone is an occupations grouping reflecting the level of education, training, and experience requirements: in Job Zone 1 occupations require little or no preparation and rely mainly on manual skills, whereas occupations in Job Zone 5 require extensive preparation and a larger share of cognitive skills.

%, depicting the interrelation between the effective AI exposure of a job, measured from the startup market signals, and its theoretical potential for exposure.
The first message we gather from Figure~\ref{fig:main} is that AIOE and AISE describe a coherent picture of AI exposure despite being  based on very different methodologies.
In the bottom left portion of the plot, where both AIOE and \startupexposure are low, we find jobs composed mainly by manual tasks, that have been identified in the literature as less subject to AI substitution \cite{felten2019occupational} and are thus less targeted by AI-based startups. In constrast, as AIOE increases, we detect a heterogeneous pattern of Occupational \startupexposure, with several occupations displaying a lower  exposure from the startup-based index with respect to AIOE.

These differences are informative about the underlying factors the two exposure metrics capture. In fact, while we learn from AIOE that some of the abilities used within a specific occupation are potentially exposed to a set of AI functions, this does not provide us information on the profitability, desirability and actual development of an AI application geared at substituting or complementing human labour. In contrast, Occupational \startupexposure takes into account already implemented, state-of-the-art AI developments, thus shedding light on a different dimension of exposure. 
To illustrate, in O*NET \textit{Database administrators} and \textit{Lawyers}  require similar sets of primarily cognitive abilities -- including, e.g., \textit{Deductive and Inductive Reasoning} or  \textit{Information Ordering}. Therefore, the two jobs present nearly identical AI potential exposure -- with AIOE scores above 6.1. However, according to our analysis, these two occupations are subject to very different degrees of exposure, with \textit{Database administrators} and \textit{Lawyers} that display an Occupational \startupexposure of about 0.8 and 0.05, respectively (AISE varies between 0 and 1). 
Needless to say, \textit{Lawyers} and \textit{Database administrations} differ for several reasons, especially linked to the societal implications of their professions. As discussed above, automating the judicial system presents both technical and ethical challenges. In contrast, there are fewer constraints on the administration of databases, a textbook example of how AI-powered tools can automate routine tasks, improve performance, potentially complementing or fully substituting human labour.
Indeed, in the region of high AIOE and low Occupational \startupexposure, there are jobs like high school teachers, judges, and marriage counsellors. Of course, AI can complement and support some secondary tasks of these jobs. However, our findings suggest that there is still no significant interest or trust in placing the essential tasks of these professions entirely in the hands of AI.

To investigate the underlying factors contributing to AI exposure, in Figure \ref{fig:main}, where each dot represents an occupation, we color occupations according to their respective Job Zones. 
In agreement with the literature on the impact of AI on the labour market, high-skilled, high-education jobs are considered the most at risk \cite{webb2019impact,acemoglu2021harms,acemoglu2022artificial}.
%,  "reserse skill-bias" with respect to the impact of computers and digital technologies 
As can be appreciated in Figure \ref{fig:main}, where higher AIOE maps into higher education and training, according to AIOE AI will disproportionately affect cognitive tasks and occupations relying mainly on problem-solving, logical reasoning, and information processing capabilities \citep{felten2021occupational}. However, our Occupational \startupexposure reveals an alternative scenario: occupations in Job Zone 5, which typically require a master's degree or higher and significant experience, are concentrated in the bottom right of the scatter plot -- with low Occupational \startupexposure and high AIOE. This indicates that despite their potential exposure to AI, as signalled by AIOE, most of these high-education and high-experience roles are not currently targeted by AI.

\begin{figure}[t!]
	\centering
	\includegraphics[width=1\textwidth]{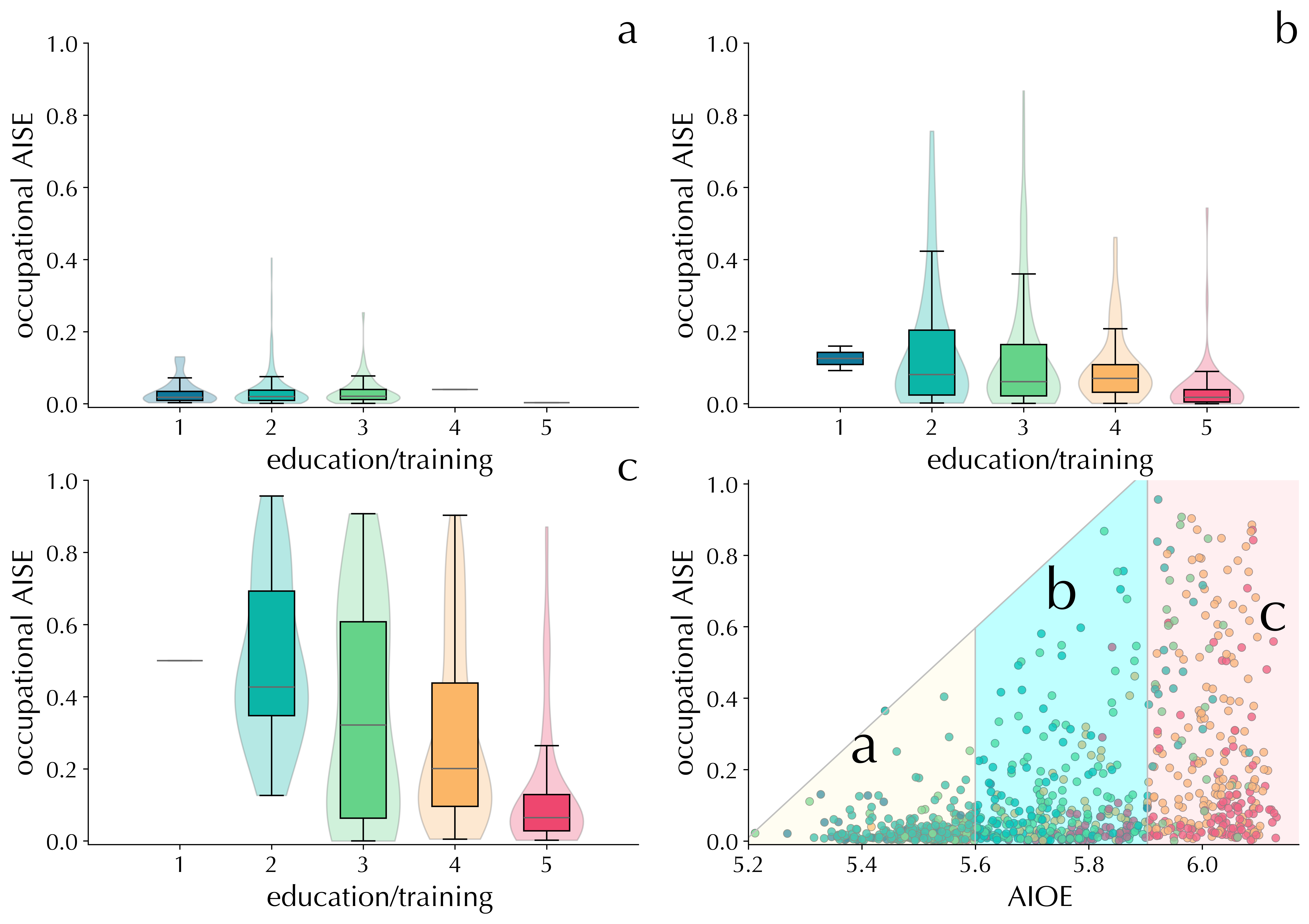}
	\caption{\textbf{\startupexposure vs Education and Training level for different fixed ranges of AIOE values.}
		Panels a-b-c: box-plots of the \startupexposure values for the different education and training levels from the O*NET \textit{job zone} feature compositions of jobs for increasing AIOE. The boxes ranges from the first quartile to the third quartile of the \startupexposure distribution of values, with a line at the median. Each whisker extends within $1.5\times$ the interquartile range. Violin plots show the distribution of the \startupexposure values extending to the possible outliers. The bottom left panel shows the regions of AIOE separation used in the previous panels divided into AIOE tertiles.}
	\label{fig:3}
\end{figure}

The separation of jobs into homogeneous intervals of AIOE highlights how the level of education and experience in an occupation relates to the expected Occupational \startupexposure, as shown in Figure  \ref{fig:3}. We divide three regions of AIOE exposure levels, in order to illustrate the key differences between \startupexposure and AIOE.
Occupations with lower AIOE aggregate in the bottom left of the \startupexposure-AIOE diagram, exhibiting uniform and low \startupexposure, as shown in panel (a) of Figure \ref{fig:3}.

In contrast, as the theoretical exposure mapped by AIOE increases, the most exposed jobs display lower education requirements, as indicated by their Job Zone.
While our analysis confirms that occupations with high potential AIOE exposure are more likely to be affected by AI, Panel (c) of Figure \ref{fig:3} reinforces our previous argument that technological feasibility is not the sole factor driving the AI-based startup market. Even with a comparable level of theoretical exposure, jobs requiring more specialized skills and education are less likely to see AI replacing humans in their essential tasks. Highly specialised professions in higher Job Zones necessitate a combination of advanced education, extensive experience, and strong cognitive, social, and human skills to handle uncertainties, such as those faced by judges or medical doctors. Complex skill sets and higher professional capacity to handle uncertainty, a trait often required in higher Job Zones, makes the practical integration of AI less straightforward, even when technically feasible.
\begin{figure}[t!]
	\centering
	\includegraphics[width=1\textwidth]{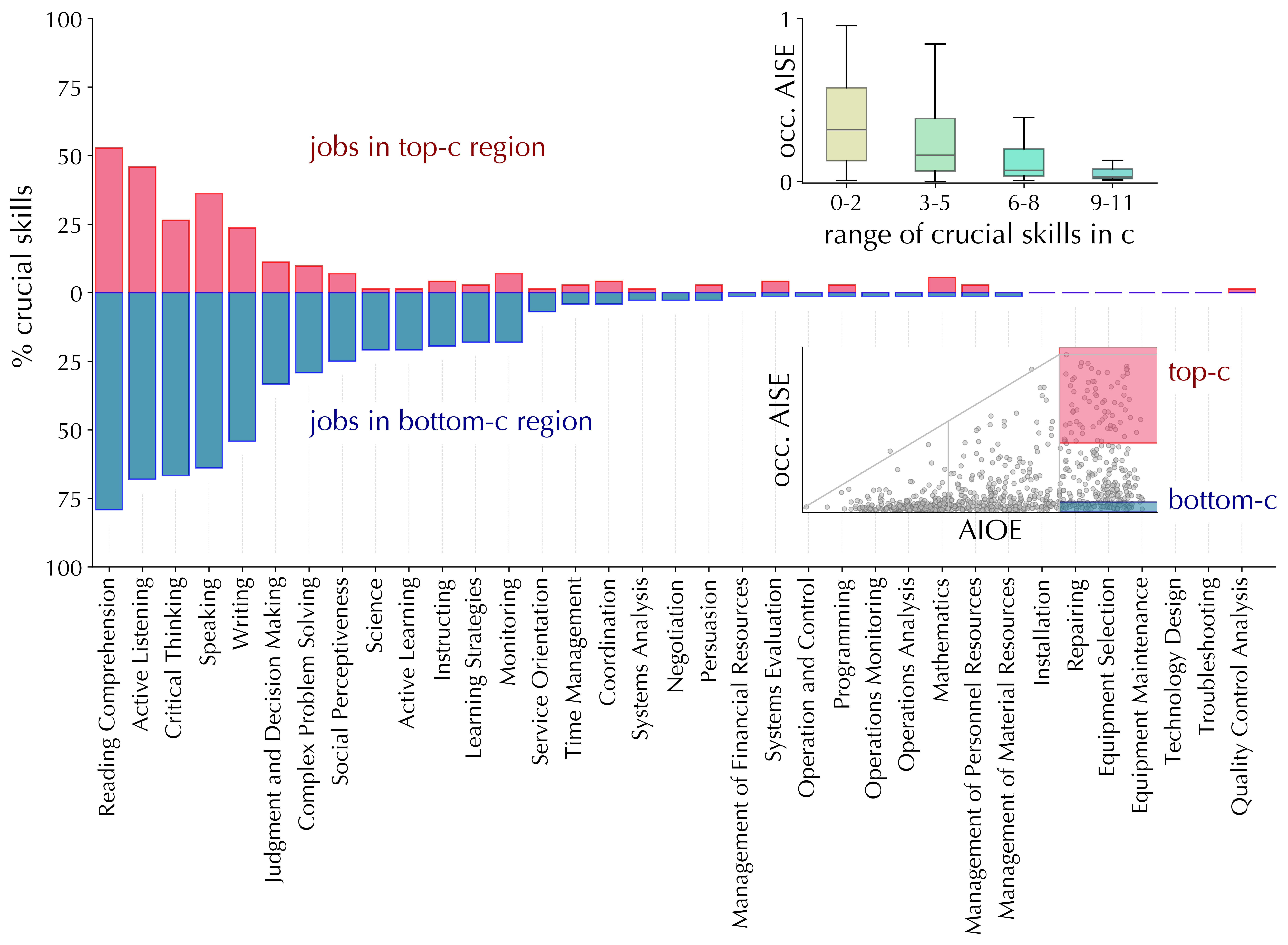}
	\caption{\textbf{Frequency of crucial skills for two regimes of the \startupexposure-AIOE diagram.}
		The bars indicate the frequency of presence of \textit{crucial skills} (skills with importance larger than 4, see the text for a detailed discussion) in the jobs in two parts of the c-region of the \startupexposure-AIOE diagram. The top red bars indicate the frequency of crucial skills in the \textit{top-c} region while the blue bars indicate the frequency of crucial skills in the \textit{bottom-c} region. The bottom inset shows the \startupexposure-AIOE diagram with the \textit{top-c} and \textit{bottom-c} regions highlighted, being the top and bottom \startupexposure quartiles of the occupations in the c-region (top AIOE tertile). The top inset shows the barplot of the \startupexposure values for different ranges of crucial skill presence in the job definitions for the jobs in the whole c-region.
	}
	\label{fig:6}
\end{figure}

%The core competency of uncertainty management is a vague and not well-defined concept.
%Whether occupations must handle complex ethical, moral, or personal tasks or constraints provided by social organizations such as laws, the database does not distinguish these tasks into skill levels. 
While in O*NET it is impossible to discern occupations requiring the capacity to handle uncertainty and complex tasks with ethical or health-related implications, we can have a better understanding of the differences between jobs with high AIOE by analyzing in detail their required skills.  O*NET associates a set of skills to each occupation, with an importance score ranging from 1 (not important) to 5 (extremely important). Crucial skills, hereafter defined as the set of skills with importance greater than 4, represent core competencies and are good indicators of an occupation's startup exposure. 
Figure~\ref{fig:6} presents a bar plot comparing the percentage of jobs with crucial skills between two areas of the scatter plot in Figure \ref{fig:main}: the upper-right (top-c) region, corresponding to occupations with high values for both AIOE and Occupational \startupexposure (blue bars) and the lower-right (bottom-c) region, containing jobs with high AIOE but low Occupational  \startupexposure (red bars). 
The first noticeable observation is that the likelihood of requiring skills with an importance score greater than 4 is significantly higher in the bottom-c region of the AIOE-Occupational \startupexposure diagram. This applies even to skills that generative AI can already manage successfully, such as \textit{reading comprehension} and \textit{writing}, and indicates that jobs in this region demand more experience or training, and errors in performing some of their associated tasks may be very costly. 

The second observation is that, in the same bottom-c region, skills related to social and human domains --such as \textit{speaking, instructing}, or \textit{judgment and decision making}-- appear more frequently. As discussed above, these skills are crucial for managing uncertainty and present challenges to AI integration, even when there is theoretical potential for AI involvement.

In summary, while there is theoretical potential for AI involvement across various occupations, the necessity for advanced skills and the high stakes associated with errors in certain jobs make AI integration less straightforward. This complexity underscores the importance of analyzing the specific skill sets required for each occupation to accurately understand their exposure to AI.
Furthermore, the presence of several crucial skills significantly influences a job's exposure to AI: having many crucial skills tends to decrease Occupational \startupexposure at fixed AIOE.
The top inset of Figure \ref{fig:6} shows the barplot splitting the range of crucial skills into four discrete intervals for jobs with high AIOE across the entire c region, as defined in Figure \ref{fig:3}.

\begin{figure}[thbp]
	\centering
	\includegraphics[width=1\textwidth]{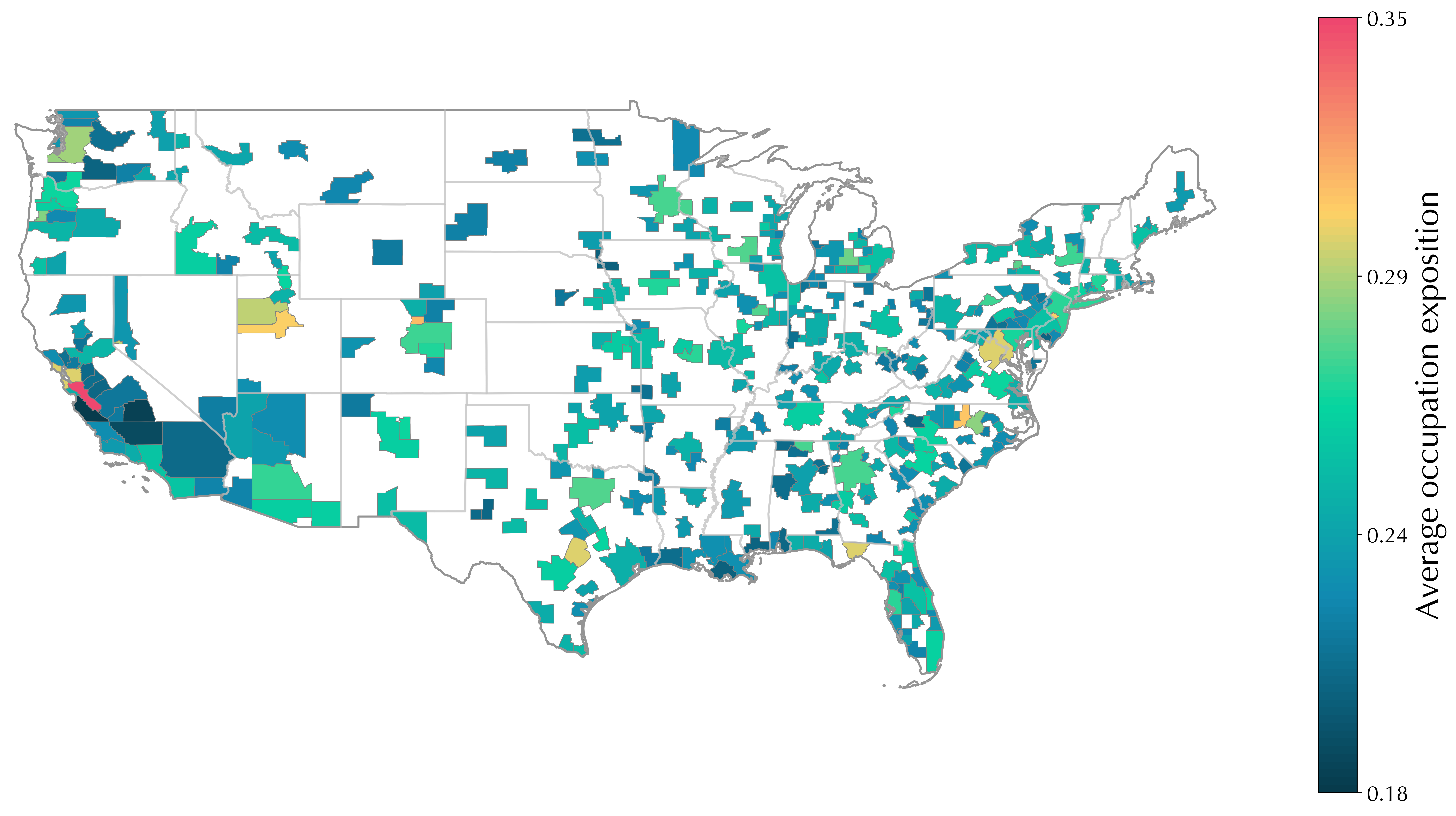}
	\caption{\textbf{Map of average exposure to AI of the workforce of different Metropolitan Statistical Areas in the USA}.
 The color-code of the figure indicates the geographical AISE of the Metropolitan Areas, that are not covering the whole nation. 
	}
	\label{fig:us_map}
\end{figure}

\subsection*{Geographical and Sectoral AI exposure}
Neither AIOE nor \startupexposure  directly measures the geographical dimension of  AI exposure for the national or sub-national workforce. However, the net exposure at the geographical level can be assessed by calculating the average occupation exposure at different geographical scales, by averaging Occupational AISE with the employment share per occupation.
Figure \ref{fig:us_map} illustrates the average Occupational \startupexposure of US Metropolitan Statistical Areas (MSAs). 
MSAs  are regions with large populations and low employment interchange with surrounding areas, thus providing a more effective representation of local labour markets than counties.
A few areas exhibit high average exposure to AI according to the \startupexposure, primarily in regions with expanding digital economies, tech industries and innovation ecosystems, such as San Jose/Santa Clara, San Francisco, Seattle, Austin, and Boulder.
Interestingly but not surprisingly, metropolitan areas in the Midwest -- encompassing Illinois, Indiana, Iowa, Kansas, Michigan, Minnesota, Missouri, Nebraska, North Dakota, Ohio, South Dakota, and Wisconsin -- display the lowest exposure to AI. This lower exposure can be attributed to the region's economic reliance on manufacturing and agriculture, industries that have been slower to integrate AI technologies compared to tech hubs. In contrast, many areas in the West show high exposure levels, aligning with the region's reputation for tech innovation and startups. This geographical disparity highlights how AI's impact on the workforce is unevenly distributed, influenced by the local economic landscape and the prevalence of digital economies.

\begin{figure}[thbp]
    \centering
    \includegraphics[width=1\textwidth]{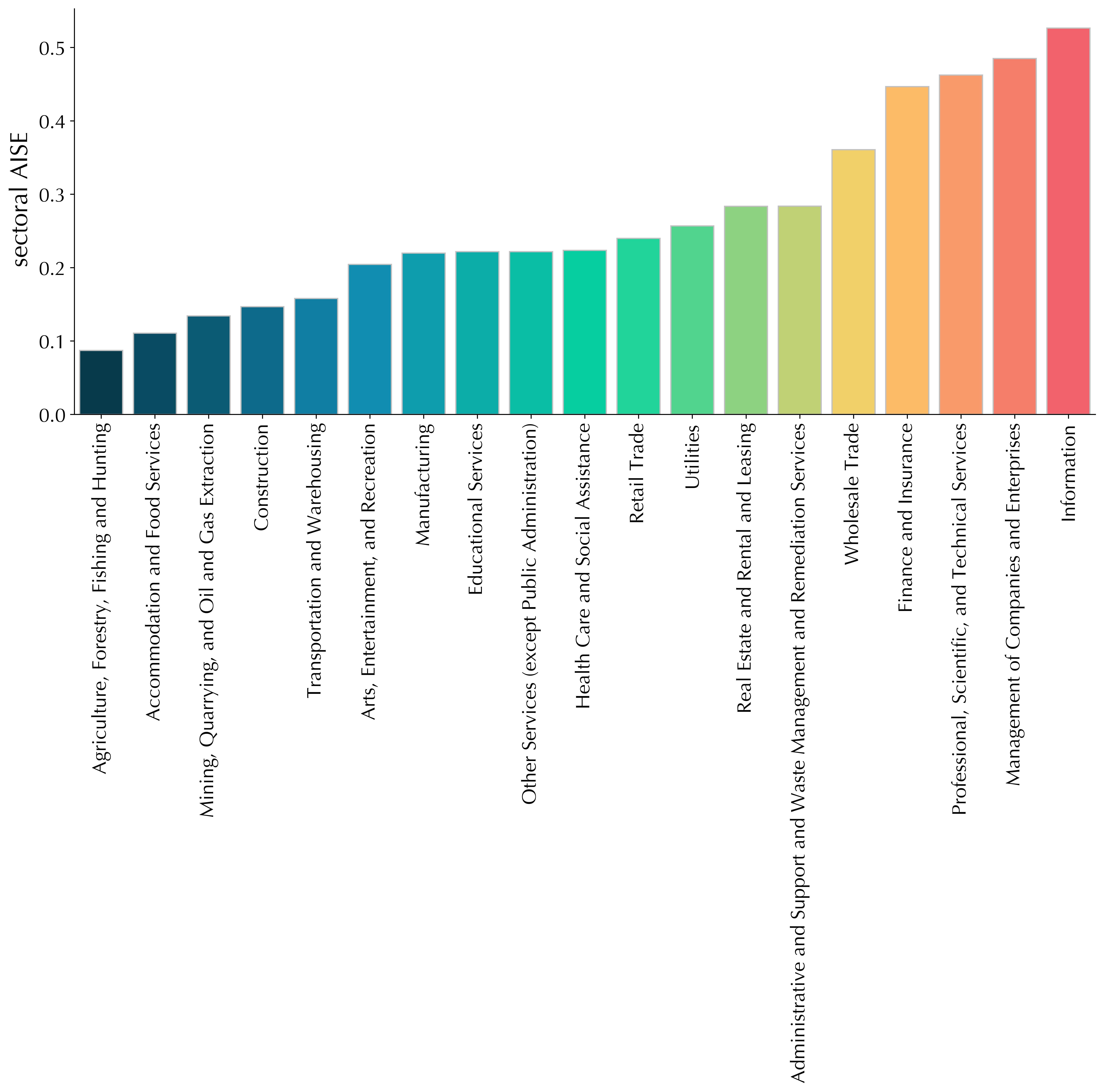}
    \caption{\textbf{Sectoral AISE} Sectors are based on two-digit NAICS classification, and occupations within sectors are weighted with the national employment data (US).
    }
    \label{fig:sectoral_aise}
\end{figure}

%In summary, while theoretical models provide insights into potential AI exposure, the practical impact on the workforce varies significantly across regions. The local economic context and the concentration of tech industries play crucial roles in determining the level of AI integration and its consequent exposure across different metropolitan areas.

%%% così?

%\begin{figure}[thbp]
%	\centering
	%\includegraphics[width=1\textwidth]{figures/fig_share_exposure_county.png}
	%\caption{\textbf{Share occupation distribution for 4 representative counties}.
	%}
	%\label{fig:county_shares}
%\end{figure}

%The results presented in this section suggest, contrary to the prevailing view, that expert professional roles will not be easily replaced by AI, at least in the short term (i.e., the time it takes for a startup to develop its potential). The confidence of society in AI is growing more slowly than the technology itself. For better or worse, this will be a brake on total automation.

We can also establish a measure of AI exposure at the sector level by combining the occupational AISE across all occupations within a given industry. Similarly to the geographical exposure, we construct a sectoral AISE index by computing a weighted average of the occupational AISE, using US Bureau of Labor Statistics industry employment levels for the nineteen two-digit NAICS sectors. 
Figure \ref{fig:sectoral_aise} displays two-digit NAICS sectors ranked in ascending order based on sectoral AISE values. Unsurprisingly, as with ranked occupations, we find that the lowest-scoring sectors are typically those involving manual L’abiura, such as construction or agriculture. Conversely, the most exposed sectors are primarily service-oriented, requiring a high level of information processing and higher levels of education (jobs in high job zones). This means they involve workers positioned in the far-right area of the AISE-AIOE diagram (zone c in Figure \ref{fig:3}d), indicating a heterogeneous level of exposure. Finally, however, sectors such as Educational Services and Health Care, which also require high education levels, show relatively low exposure for the same reasons mentioned earlier, as they include many jobs located in the lower-right area of the AISE-AIOE diagram.

\section*{Discussion}

This study introduces the AI Startup Exposure (AISE) index as a novel tool to assess the impact of artificial intelligence (AI) on occupations, industries, and geographic areas, using real-world data from venture-backed startups. Unlike AI exposure indices based on patents, AI benchmarks, and crowd-sourced evaluations, which often fail to capture real-world adoption and focus on technological feasibility without considering economic viability, AISE focuses on tangible innovations developed by AI startups funded by Y Combinator.
This gives the index a future-oriented and economically grounded perspective, while avoiding the limitations of existing indices of potential exposure, such as crowd-sourced assessments that can lead to speculative predictions, and patent-based measures that suffer from both intrinsic time lags in data collection and the fundamental issue that many AI applications are not patented or patentable.

Our analysis reveals a significant finding: while existing indices of potential exposure suggest that high-skilled, highly educated occupations—such as judges, pediatricians, and other expert-driven roles—are the most exposed to AI, the AISE index paints a different picture. Startups are not primarily targeting these professions for automation likely due to technical, societal, and ethical constraints, particularly in high-stakes domains. In contrast, occupations with routine tasks, such as office clerks, and data analytics tasks, such as data scientists and market research analysts, show much higher AI exposure.%, as startups focus on automating tasks that are economically viable and technically feasible.
The sectoral analysis reveals a heterogeneous distribution of AI’s impact across industries,  with high-tech and finance sectors exhibiting higher exposure to AI, while traditional industries such as agriculture and manufacturing demonstrate comparatively lower exposure levels. The geographic analysis highlights that areas with knowledge intensive economies, particularly San Francisco and Seattle, face markedly higher AI exposure, whereas the Midwest, characterized by its manufacturing base, exhibits substantially lower exposure.
Additionally, this study explores the integration of AI with robotics, an emerging area with significant potential for workforce disruption, by targeting the products and services developed by AI-robotics startups. Our Startup Robotic Exposure (RSE) index indicates that many occupations with low AI exposure, particularly those involving manual tasks or physical abilities, are more exposed to robotics. This is especially true for lower-skill jobs, where robotics and AI are being integrated to automate physical tasks. However, many high-exposure AI jobs are also increasingly integrated with robotics, suggesting that the joint action of AI and robotics could transform occupations beyond manufacturing, including clerical and information-processing roles. These preliminary findings highlight the need for further research on two fronts: first, to better understand how the convergence of AI and robotics could amplify technological disruption across sectors, and second, to expand the data sample beyond the current limited number of AI-robotics startups.

A significant advantage of our exposure index is its ability to be continuously refined and expanded as more AI startups emerge. The index can be easily updated with larger datasets from other startup ecosystems beyond Y Combinator, providing a more comprehensive view of AI innovation. Additionally, combining AISE with other sources of information, such as AI patents and scientific publications, would offer an even broader and more nuanced understanding of AI’s impact. 

While the AISE and RSE indices offer valuable insights, several limitations warrant consideration. The reliance on startup data introduces a bias towards innovations that are already commercially viable or secured venture capital backing, potentially underestimating AI developments in academia or industries that are less reliant on venture capital. Moreover, focusing on the occupational dimension, our index, by design, operates the implicit assumption that occupations are homogeneous across geography and across firms. However this is not always the case, occupational heterogeneity can stem from the significant portion of AI-related tasks driven by the tacit knowledge embedded in firm-specific organizational practices and procedural routines \citep{dosi2015dynamics,dosi2019whither}, or from different labour market contexts \citep{de2023creates}. This makes AI exposure far more complex than occupational-level analysis can reflect \citep{tolan2021measuring}. To address these firm-level dynamics, a more granular approach beyond occupational-level exposure is necessary for a comprehensive understanding of AI's employment effects.

In conclusion, this study offers a new perspective on AI's labour market implications by grounding its analysis in market-driven innovations rather than theoretical models. Contrary to widespread fears of imminent job displacement in high-skilled occupations, the AISE index suggests that automation will target routine, economically viable tasks first, while expert-driven roles will remain relatively shielded from AI disruption at least in the short term. Moreover, by combining startup activity with real-time tracking of AI developments, the AISE index can inform more effective policymaking and workforce planning, offering a dynamic tool to monitor AI's evolving impact.

Looking ahead, while this study remains deliberately silent on the interpretation of geographical and sectoral exposure patterns, future research will investigate the labour market implications of our exposure indices. Specifically, we plan to examine how differential AI exposure across regions and sectors translates into wage dynamics, employment patterns, and inequality outcomes. Furthermore, we aim to enhance the robustness of our occupational analysis by incorporating alternative occupational and task descriptions, e.g. more relevant to the European context or to developing economies’ labour markets, moving beyond the US-centric O*NET database. Additionally, the continuous refinement of the AISE index, incorporating larger databases and diverse textual sources of AI advancements, will provide policymakers with an effective tool to better anticipate labour market trends and develop informed strategies to address the challenges and opportunities presented by this new wave of technological change.

\section*{Methods}  \label{sec:data&methods}

\subsection*{Data}
\label{sec:data}
\subsubsection*{Occupational data}
To characterise occupations we rely on O*NET (United States Occupational Information Network)\footnote{O*NET is publicly available at: \url{www.onetcenter.org}} maintained by the US Department of Labor's Employment and Training Administration (ETA). 

O*NET provides survey-based information about the skills, knowledge, tasks, tools, technologies, and educational requirements connected to each occupation, organized according to the O*NET-SOC classification \citep{gregory2019updating}. In our analysis, we employ the O*NET-SOC lowest level of aggregation, thereby extracting information on 1016 jobs and rely on three sets of occupation-level information: tasks, skills, and education and training requirements. 

First, and crucial to our methodology, to connect jobs to start-up AI advancements, we consider each occupation's summary description, i.e., a summary of its essential tasks. We purposely use these short description rather then the full task lists, because they report effectively the salient characteristics of the job while being concise, an essential feature for our analysis, since we use these descriptions as input for our queries to the large language model Llama 3. For instance, the job description for Cardiologists reads: "Diagnose, treat, manage, and prevent diseases or conditions of the cardiovascular system. May further subspecialize in interventional procedures (e.g., balloon angioplasty and stent placement), echo-cardiography, or electrophysiology". Feeding Llama3 with such a job's description that summarizes essential tasks in two or three sentences as opposed to the entire set of individual job's tasks has the advantage of reducing noise because the complete task lists of a SOC occupation can be extensive and can include detailed, context-specific tasks that are not central to the primary role of the job. Therefore, employing the O*NET short description reduces the risk of overloading the language model with excessive detail while maintaining sufficient information about the essential features of jobs.

Second, we use occupational skills. O*NET defines a set of 35 skills --such as \textit{writing, reading comprehension}, or \textit{coordination}, that are associated to each occupation with an importance score ranging from 1 (not necessary) to 5 (essential). We use this importance score to analyze job characteristics in Figure \ref{fig:6}.

Third, we retrieve information on educational, experience and on-the-job training requirements as described by O*NET Job Zones, that group occupations based on their similarity in human capital requirements. For instance, a job in Job Zone 1 requires little or no preparation, while a job in Job Zone 5 requires extensive preparation -- more details on Job Zones can be found at: \url{www.onetonline.org/help/online/zones}. %In particular, we use the Job Zones to characterize jobs in Figures \ref{fig:main}, \ref{fig:2}, and \ref{fig:3}.

Finally, to plot the maps in Figures \ref{fig:us_map} and \ref{fig:us_map_1}, we obtain data about employment level for SOC occupations from the \textit{Quarterly Census of Employment and Wages} (QCEW) dataset of the US \textit{Bureau of Labor Statistics} (\url{https://www.bls.gov/cew/}). 

\subsubsection*{AI \& Robotic startup data}
\begin{figure}[htbp]
    \centering
    \includegraphics[width=1\textwidth]{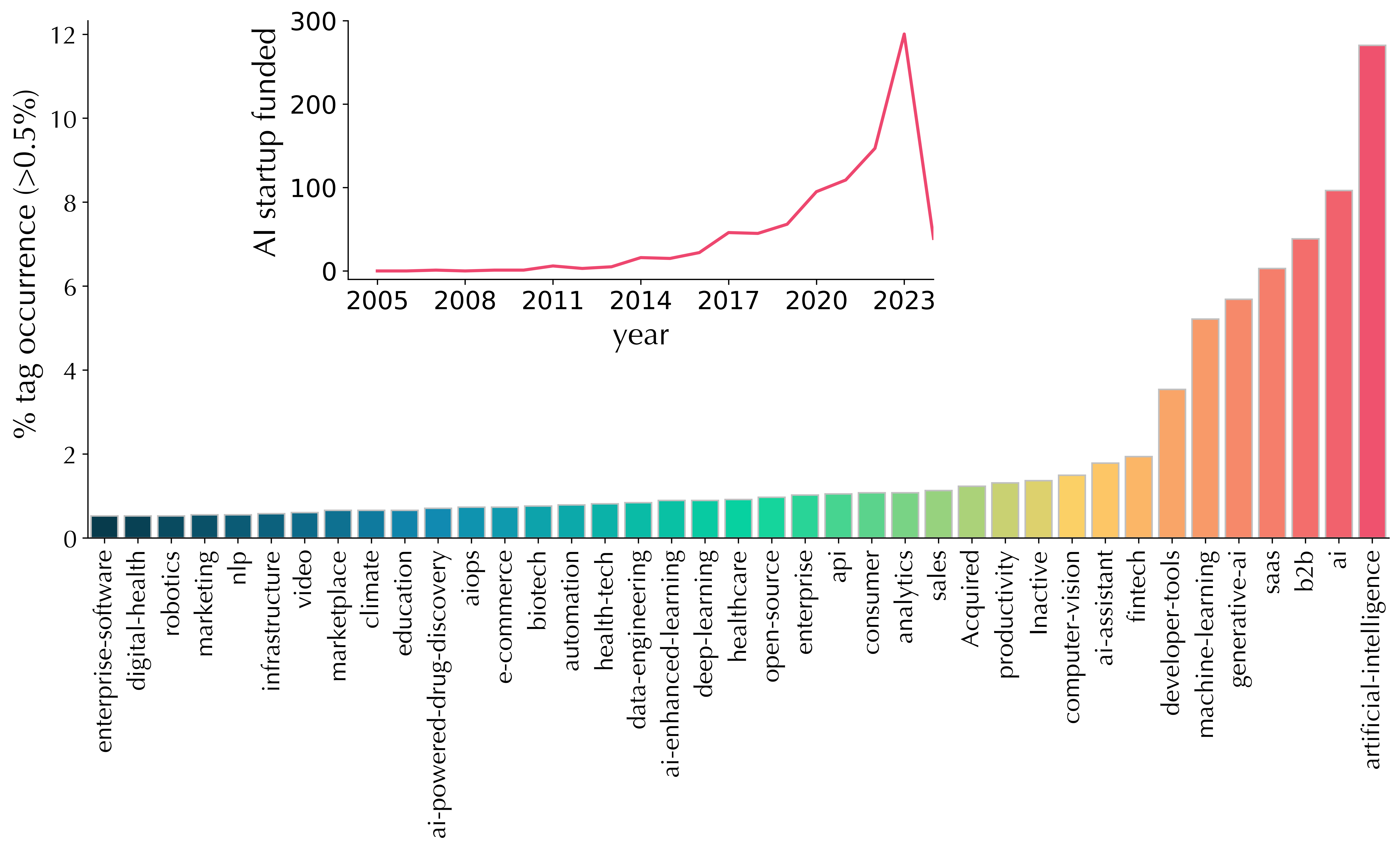}
    \caption{\textbf{Histograms of the frequency of tags among the AI-tagged startups}.
    The inset shows the number of AI-tagged startups over time.}
    \label{fig:AI_vs_Time}
\end{figure}

To obtain information on AI innovative startups we rely on the set of startups that were financed by the US-based technology startup accelerator and venture capital firm Y Combinator (YC), by web scraping YC's website \url{www.ycombinator.com/companies}. 
The website contains information on over 5000 startups funded since YC's launch in 2005, all of which have received a fixed amount of \$500000.
In particular, for each startup, we extract the name, a brief and a detailed description, the YC funding year, and a set of thematic tags defined by YC. To construct the \startupexposure, we select only the startups with AI-related tags and use their detailed descriptions as input for queries to Llama 3. More specifically, we consider the following tags: \textit{AI, artificial intelligence, AI-assistant, AI-powered drug discovery, AIOps, conversational AI, ML, machine learning, deep learning, deepfake detection, generative AI, AI-enhanced learning} and \textit{ computer vision}. Figure \ref{fig:AI_vs_Time} shows the frequency of all tags among the selected startups.% reinforcement learning}, and \textit{speech recognition}.
We thus obtain a subset of 958 AI startups, funded by YC between 2005 and March 2024. However, as shown in the inset of Figure \ref{fig:AI_vs_Time}, more than $50\%$ of the AI startups was funded after 2020.

Similarly, to construct the RSE index, we fed Llama3 with the descriptions of startups that had a robotics-related tag, and we considered the following tags: \textit{Robotics, Robotic Process Automation, Food Service Robots \& Machines, Medical Robotics}, and \textit{Robotic Surgery}, extracting 103 startups. Note that the number of startups with these tags is limited to just over a hundred because most of the robotics startups funded by YC are relatively new and focus on integrating robotics and AI. Since the integration of AI in robotics is an emerging field, the results on the exposure of professions to robotics presented in this paper are preliminary and will be explored further in future works. 

Other than the YC tags, the startup description are unstructured, e.g., the platform \textit{Dili} is described as follows: "Dili is a due diligence co-pilot platform that leverages AI on public and private datasets to help investors make better investment decisions with generative insights and analysis. Using Dili, Investors can enhance and automate their due diligence workflows such as automatically screening deals, generating comps analysis, finding deal killers, managing the due diligence request list process, search across their firm’s entire corpus of deals, and more".

%\begin{itemize}
    %\item Y-combinator AI-tagged start-up data-set
    %\subitem AI tagged start-up plot
 %  \item LLM approach to quantify exposure: construction of the Y-combinator-based forward-looking exposure index linking job tasks to new AI developments
%\end{itemize}
\subsection*{LLM approach to quantify AI startup exposure}
While not all Y Combinator-funded startups have been successful or are still active, they have all secured funding and proposed an innovative AI application, an important signal in detecting the potential exposure of jobs to recent advancements in artificial intelligence. Therefore, to construct our AI occupational exposure index, we select all the description of AI-tagged startups funded by Y Combinator and feed them to a large language model, as mentioned above here we rely on Llama 3, to infer if they developed/are developing a product or a service with the potential to replace some or all tasks associated to a job. For efficient and reproducible execution, we use the pre-trained 8B version of Llama 3, i.e. the open-weight LLM released by Meta in April 2024 (see \url{https://ai.meta.com/blog/meta-llama-3/}). For each O*NET-SOC occupation, we iterate over all AI-tagged startups and feed the following prompt to Llama 3:\\
\\
$\{$\textit{“role”: “system”, “content”: “You are an AI specialist.”}$\}$ \\
$\{$\textit{“role”: “user”, “content”: “Given the following startup description: ” + startup[j] + “and given the following job description: “+ job[i] + “can the product or service developed by the startup directly replace humans to perform some of the described job’s tasks? Use only the information provided by the two descriptions. Reply only yes or no.”}$\}$\\
\\
Where startup[j] represents the detailed description of startup j, and job[i] the summary of essential tasks necessary to perform job i -- as provided by O*NET. Each occupation's AI exposure  is then computed as the sum of startups for which Llama 3 responded \textit{yes} normalized by the number of AI tagged startups.

It is important to bear in mind that even though our prompt asks to the LLM whether the startup can replace some essential tasks of the job, our measure is not a measure of substitutability. In fact,  to get an affirmative (\textit{yes}) reply to our prompt, it is sufficient that a single task is deemed replaceable by Llama 3. Therefore, our index  does not aim at examining the potential labour saving impact of AI, in contrast, it should be interpreted as a measure of occupational AI exposure index as measured by the interest of the startup ecosystem in broadly influencing/interacting with a particular job. 

To test the robustness of our methodology and results, we repeat the above experiment in two different settings. In the first case, we use the same prompt described above but employed the short instead of the detailed descriptions of the startups. The detailed descriptions often provide additional information unrelated to the product or service developed by the startup -- i.e. about the founders or the investments received -- which may potentially be a source of noise. In contrast, the short descriptions are less detailed, but focus solely on the product being developed, reducing the risk of Llama3 being confused by irrelevant details. As can be appreciated in the SI, 
however, the results obtained using the short descriptions are highly consistent with those obtained using the detailed descriptions.

As an additional test, we also modify the prompt as follows:\\
\\
$\{$\textit{“role”: “system”, “content”: “You are an AI specialist.”}$\}$ \\
$\{$\textit{“role”: “user”, “content”: “Given the following startup description: ” + startup[j] + “and given the following job description: “+ job[i] + “Is the product or service developed by the startup designed to directly replace humans to perform some of the described job’s tasks? Use only the information provided in the two descriptions. Reply only with yes or no.”}$\}$\\
\\
The question posed to Llama 3 in this prompt is more direct, as it explicitly asks whether the startup is developing a product specifically intended to replace one or more job tasks, while the previous prompt allowed Llama 3 greater freedom in interpreting the connection between the startup's AI product/service and the job's tasks. For instance, a startup developing self-driving cars may not explicitly state that it also produces self-driving school buses. Using the question in the second prompt, Llama 3 would infer that school bus drivers are not exposed to the startup's AI application, whereas using the first prompt it would. Nevertheless, the exposure rankings obtained with the two different prompts are highly consistent (see SI for more details).

Finally, we have to mention that our approach has some inherent limitations that leave room for improvement. First, while Y Combinator is a highly representative and reliable source in the international ecosystem of AI-related startups, it is certainly not exhaustive, and there may be biases in selecting and funding certain types of startup applications over others. For example, from Fig.
ef{fig}, we observe that the tag "generative AI" is significantly more prevalent than the tag "computer vision." On one hand, this indicates greater interest in generative AI applications (which is what our AISE index measures). However, we cannot determine whether this difference is also due to a bias at Y Combinator, which, for several reasons (e.g., internal policy), may favor selecting these types of applications. Therefore, future works should aim to collect more startup datasets from different funding sources. However, unlike other startup datasets, Y combinator is freely available and well-structured. Secondly, Llama 3 is not free from noise, and our methodology could benefit from using more advanced language models in the future.% Therefore, the value of our \startupexposure index for a job should be interpreted as relative to other jobs, just as other existing technological indicators in the literature.

\subsection*{AI Occupational Exposure (AIOE) index}
To test and validate our measure of startup-based AI exposure, we compare Occupational AISE with a standard measure in the literature, the AI Occupational Exposure (AIOE) index introduced by Felten et al. \cite{felten2021occupational}. AIOE connects ten AI applications sourced from the Electronic Frontier Foundation AI Progress Measurement project, such as image recognition or text generation, with 52 O*NET occupational abilities, such as \textit{oral comprehension} and \textit{inductive reasoning}. The AI application-ability degree of relatedness is established through a matrix crowd-sourced from Amazon's Mechanical Turk provided in \cite{felten2021occupational}, and an ability exposure is defined as the sum of its relatedness scores with the AI applications. The AIOE of each occupation is then computed as the weighted average of the exposures of the abilities used within the occupation. The weights are derived from O*NET ability \textit{level} (ranging from 1 to 7) which indicates the degree to which an ability is required to perform the job tasks, and the ability \textit{importance} (ranging from 1 to 5) which indicates how critical the ability is for performing the job tasks. Since in O*NET, the job characteristics are periodically updated, to build the AIOE measure employed in this paper we draw the ability level and importance from O*NET 2024, that slightly differ from the version of O*NET used in \cite{felten2021occupational}. Therefore, the AIOE for each job $i$ is computed as follows:

\[\text{AIOE}_i= \frac{\sum_{j=1}^{52} A_{kj} L_{ij} I_{ij}}{\sum_{j=1}^{52} L_{ij} I_{ij}}\]

where $k$ represents the AI application, $j$ the ability, and $A_{kj}$ the exposure to AI of ability $j$. O*NET provides occupation-ability connections for 873 jobs. Therefore, we compare our exposure index with the AIOE only for these 873 jobs, a subset of the 1016 jobs we analyze with the startup data and Llama 3.

AIOE, not only bears methodologically different from our AISE measure as it is survey-based rather than LLM-based, but is also conceptually distinct. While our index calculates the actual exposure of a job to AI-related startups directly from O*NET occupation description, AIOE is a ability-based measure \cite{tolan2020}, i.e., is a weighted sum of ability-level exposures scores.  %i.e., firstly the ability exposure is assessed and secondly aggregates these skills at the occupational level. 
By identifying which abilities are more or less likely to be performed by AI \cite{staneva2023measuring} and considering occupations not as a whole but as aggregations of a bundle of abilities with different exposure levels, AIOE is a forward looking measure of the "potential" rather than actual impact of AI on jobs \cite{felten2021occupational}. This means that a job requiring many abilities highly exposed to AI is potentially exposed as other factors, not explicitly covered by the sole ability dimension, may be at play and may mitigate or worsen its exposure, as extensively argued in the results section.
Moreover, upon inspecting the abilities in O*NET, we note that they are not highly granular when it comes to cognitive abilities. For example, the ability \textit{Written Expression} encompasses various writing skills, some of which—like summarizing texts and writing reports—are already mastered at high levels by AI (particularly by modern language models), while others, such as creative writing, are areas where AI still lags behind. As a result, jobs that require this type of writing are currently less exposed to automation. This further highlights that the AIOE measures potential exposure rather than actual exposure, since the less granular abilities used are too general to refer to specific job tasks.

\section*{Acknowledgements}
D.M., A.P., A.S., and A.T. acknowledge the financial support under the National Recovery and Resilience Plan (NRRP), Mission 4, Component 2, Investment 1.1, Call for tender No. 1409 published on 14.9.2022 by the Italian Ministry of University and Research (MUR), funded by the European Union – Next Generation EU – Project Title "Triple T – Tackling a just Twin Transition: a complexity approach to the geography of capabilities, labour markets and inequalities" – CUP F53D23010800001 - Grant Assignment Decree No. 1378 adopted on 01.09.2023 by the Italian Ministry of Ministry of University and Research (MUR).

\bibliographystyle{agsm}
%\newpage
\bibliography{bibliography.bib}

\clearpage

\section*{Supplementary Information}
\tableofcontents

\sisubsection{Most and least exposed jobs according to the Occupational AISE index}
In the tables below, we show the most and least exposed jobs to startups AI applications, according to our \startupexposure index.

As emphasized in the main text, jobs most exposed to AI startup applications share a common set of tasks. These roles predominantly involve processing and analyzing information, often in structured formats, along with planning and organization. They typically do not demand significant social interaction but frequently require computer and programming skills. Consequently, these tasks often involve working with a monitor as an essential tool for mediation.

On the other hand, the less exposed jobs and their associated tasks fall into more diverse categories, such as medical specialists, technical workers, and religious officials. These positions are not the primary targets of startups for various reasons, many of which are quantitatively analyzed in the main text. For instance, these roles require complex manual skills, frequent social interaction and emotional intelligence, and often involve dealing with ethical and moral issues or with high-risk situations where even a single error can have serious consequences

The case of the airline pilot is particularly noteworthy. One might assume that this role is highly exposed to AI, given that modern airplanes already incorporate AI for all phases of flight. However, it is crucial to remember that the \startupexposure index derives its insights from the most recent AI-based startups. Consequently, it is not necessarily evident that new startups would aim to automate tasks that have long been managed by existing technologies.
Moreover, our index seeks to shed light on the influence of AI on current jobs. Given that AI is already integrated into the responsibilities of an airline pilot, the low \startupexposure value for this profession suggests that there is little interest (or feasible efforts) in developing new technologies beyond those already in place to further automate the role of an airline pilot.

\begin{table}[ht]
   \centering
    % Place the caption above the table
    \caption{\textbf{Most exposed jobs according to the \startupexposure index.\\
    \\}}
    \label{tab:1}
    \begin{tabularx}{\textwidth}{|p{3cm}|X|}
        \hline
        \textbf{Job title} & \textbf{Job description (O*NET)} \\
        \hline
        \textit{Office Clerks, General} & Perform duties too varied and diverse to be classified in any specific office clerical occupation, requiring knowledge of office systems and procedures. Clerical duties may be assigned in accordance with the office procedures of individual establishments and may include a combination of answering telephones, bookkeeping, typing or word processing, office machine operation, and filing. \\
        \hline
         \textit{Data scientist} & Develop and implement a set of techniques or analytics applications to transform raw data into meaningful information using data-oriented programming languages and visualization software. Apply data mining, data modeling, natural language processing, and machine learning to extract and analyze information from large structured and unstructured datasets. Visualize, interpret, and report data findings. May create dynamic data reports. \\
        \hline
        \textit{Interviewers, Except Eligibility and Loan} & Interview persons by telephone, mail, in person, or by other means for the purpose of completing forms, applications, or questionnaires. Ask specific questions, record answers, and assist persons with completing form. May sort, classify, and file forms. \\
        \hline
        \textit{Computer and Information Systems Managers} & Plan, direct, or coordinate activities in such fields as electronic data processing, information systems, systems analysis, and computer programming. \\
        \hline
        \textit{Executive Secretaries and Executive Administrative Assistants} & Provide high-level administrative support by conducting research, preparing statistical reports, and handling information requests, as well as performing routine administrative functions such as preparing correspondence, receiving visitors, arranging conference calls, and scheduling meetings. May also train and supervise lower-level clerical staff.\\
        \hline
        \textit{Market Research Analysts and Marketing Specialists} & Research conditions in local, regional, national, or online markets. Gather information to determine potential sales of a product or service, or plan a marketing or advertising campaign. May gather information on competitors, prices, sales, and methods of marketing and distribution. May employ search marketing tactics, analyze web metrics, and develop recommendations to increase search engine ranking and visibility to target markets.
        \\

        \hline
    \end{tabularx}
    %\caption{6 most exposed jobs according to the \startupexposure index.}
    %\label{tab:sample}
\end{table}

\begin{table}[ht]
   \centering
    % Place the caption above the table
    \caption{\textbf{Least exposed jobs according to the \startupexposure index.\\
    \\}}
    \label{tab:2}
    \begin{tabularx}{\textwidth}{|p{5cm}|X|}
        \hline
        \textbf{Job title} & \textbf{Job description (O*NET)} \\
        \hline
        \textit{Clergy} & Conduct religious worship and perform other spiritual functions associated with beliefs and practices of religious faith or denomination. Provide spiritual and moral guidance and assistance to members. \\
        \hline
         \textit{Acupuncturist} & Diagnose, treat, and prevent disorders by stimulating specific acupuncture points within the body using acupuncture needles. May also use cups, nutritional supplements, therapeutic massage, acupressure, and other alternative health therapies. \\
        \hline
        \textit{Pediatric Surgeons} & Diagnose and perform surgery to treat fetal abnormalities and birth defects, diseases, and injuries in fetuses, premature and newborn infants, children, and adolescents. Includes all pediatric surgical specialties and subspecialties.

 \\
        \hline
        \textit{Terrazzo Workers and Finishers} & Apply a mixture of cement, sand, pigment, or marble chips to floors, stairways, and cabinet fixtures to fashion durable and decorative surfaces. \\
        \hline
        \textit{Nurse Anesthetists} & Administer anesthesia, monitor patient's vital signs, and oversee patient recovery from anesthesia. May assist anesthesiologists, surgeons, other physicians, or dentists. Must be registered nurses who have specialized graduate education.\\
        \hline
        \textit{Airline Pilots, Copilots, and Flight Engineers} & Pilot and navigate the flight of fixed-wing aircraft, usually on scheduled air carrier routes, for the transport of passengers and cargo. Requires Federal Air Transport certificate and rating for specific aircraft type used. Includes regional, national, and international airline pilots and flight instructors of airline pilots.
        \\

        \hline
    \end{tabularx}
    %\caption{6 most exposed jobs according to the \startupexposure index.}
    %\label{tab:sample}
\end{table}

\clearpage
\sisubsection{Exposure indices vs job Zones}
In the main text, we illustrated how the \startupexposure index varies across job zones within fixed intervals of AIOE. Our findings indicate that, for a given AIOE interval, jobs that demand higher levels of education and training are less likely to be targeted by AI-based startups. The figure below displays the overall relationship between job zones and the \startupexposure, independent of AIOE. For comparison, we also present the general relationship between job zones and AIOE.

The figure shows that both indices increase from job zone 1 to job zone 4. The prevalence of manual L’abiura in the lower job zones complicates the application of AI. However, only our \startupexposure index exhibits a notable decline between job zone 4 and job zone 5. This difference arises because, unlike the AIOE, which is an ability-based index that considers jobs requiring more cognitive skills as more exposed, our \startupexposure index reflects the genuine interest of the startup market in AI applications. As highlighted in the main text, jobs demanding the highest levels of education and training often entail significant ethical and social considerations and involve higher-risk scenarios, making the integration of AI less straightforward compared to most jobs in job zone 4

\begin{figure}[htbp]
    \centering
    \includegraphics[width=1\textwidth]{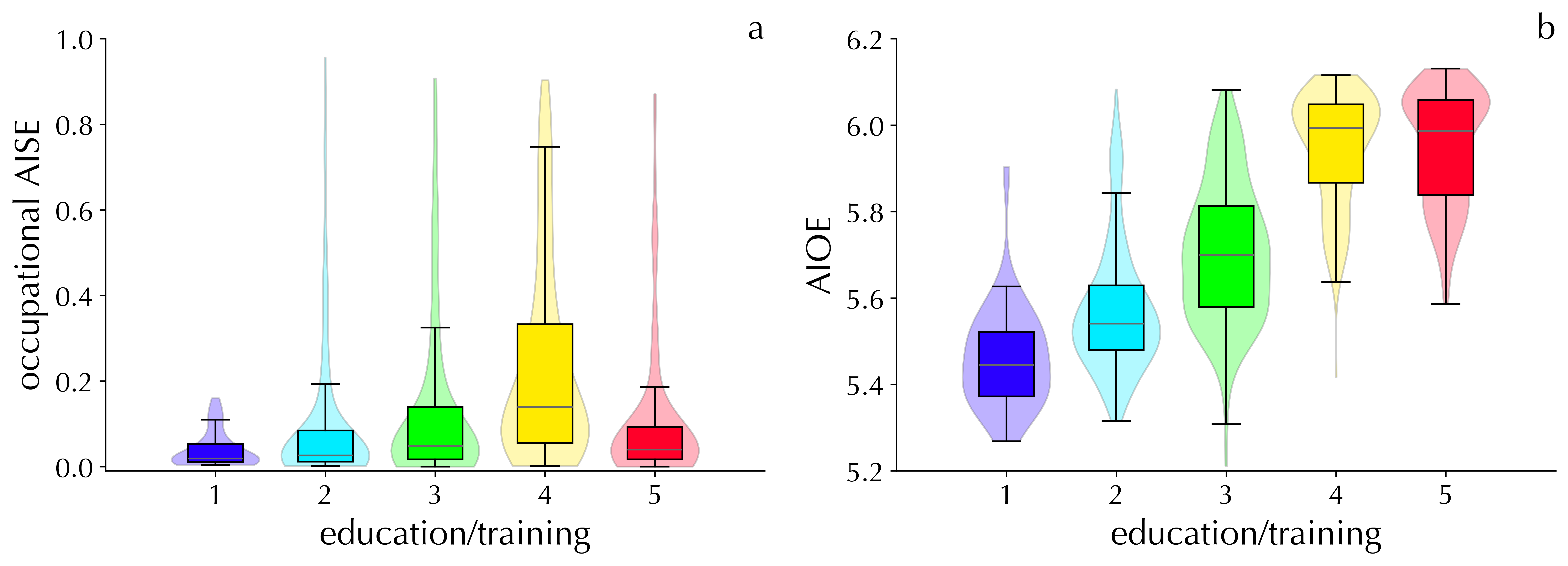}
    \caption{\textbf{\startupexposure and AIOE vs Job zone.} The width of the violin is proportional to density of jobs within the corresponding interval of exposition values and the black dot represents the average. The number of points for each violin is denoted by $n$.}
    \label{fig:2}
\end{figure}
\clearpage

\clearpage
\sisubsection{Exposure indices vs number of \textit{crucial} skills}
The two figures below illustrate how the average AIOE and \startupexposure for jobs vary with the number of skills required that have an O*NET importance score greater than 4. Consistent with the literature suggesting that high-skilled jobs are the most exposed, we observe that the AIOE increases with the number of skills required that have an importance score above 4. However, for jobs within a fixed AIOE range, the \startupexposure index decreases as the number of required skills with an importance score greater than 4 increases.

This reflects that, given a fixed theoretical automation potential (AIOE), there is a preference—especially from Y Combinator startups—to invest in AI applications targeting the automation of less complex (low-skilled) tasks. Consequently, there is less confidence in integrating AI for tasks that require numerous critical skills to handle risky or uncertain situations.

\begin{figure}[htbp]
    \centering
    \includegraphics[width=0.8\textwidth]{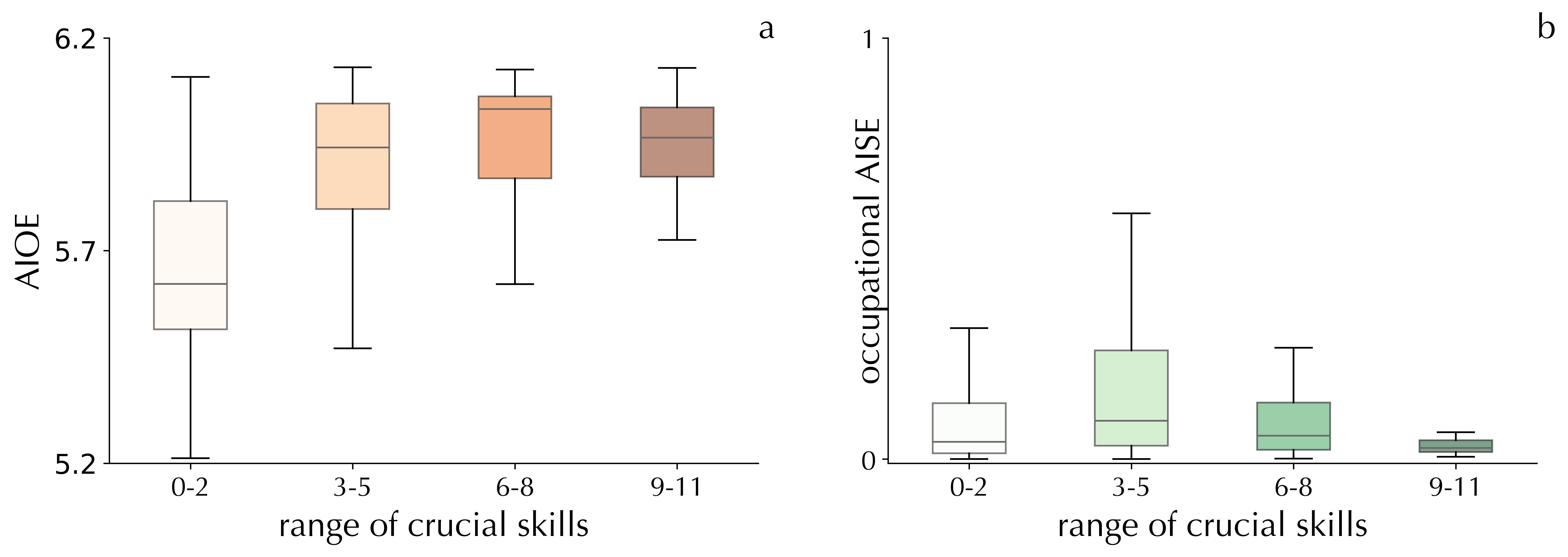}
    \caption{\textbf{AIOE vs number of skills with importance larger than 4. for different fixed ranges of AIOE values.} The width of the violin is proportional to density of jobs within the corresponding interval of Y Combinator values and the black dot represents the average. The number of points for each violin is denoted by $n$.}
    \label{fig:4}
\end{figure}

\begin{figure}[htbp]
    \centering
    \includegraphics[width=1\textwidth]{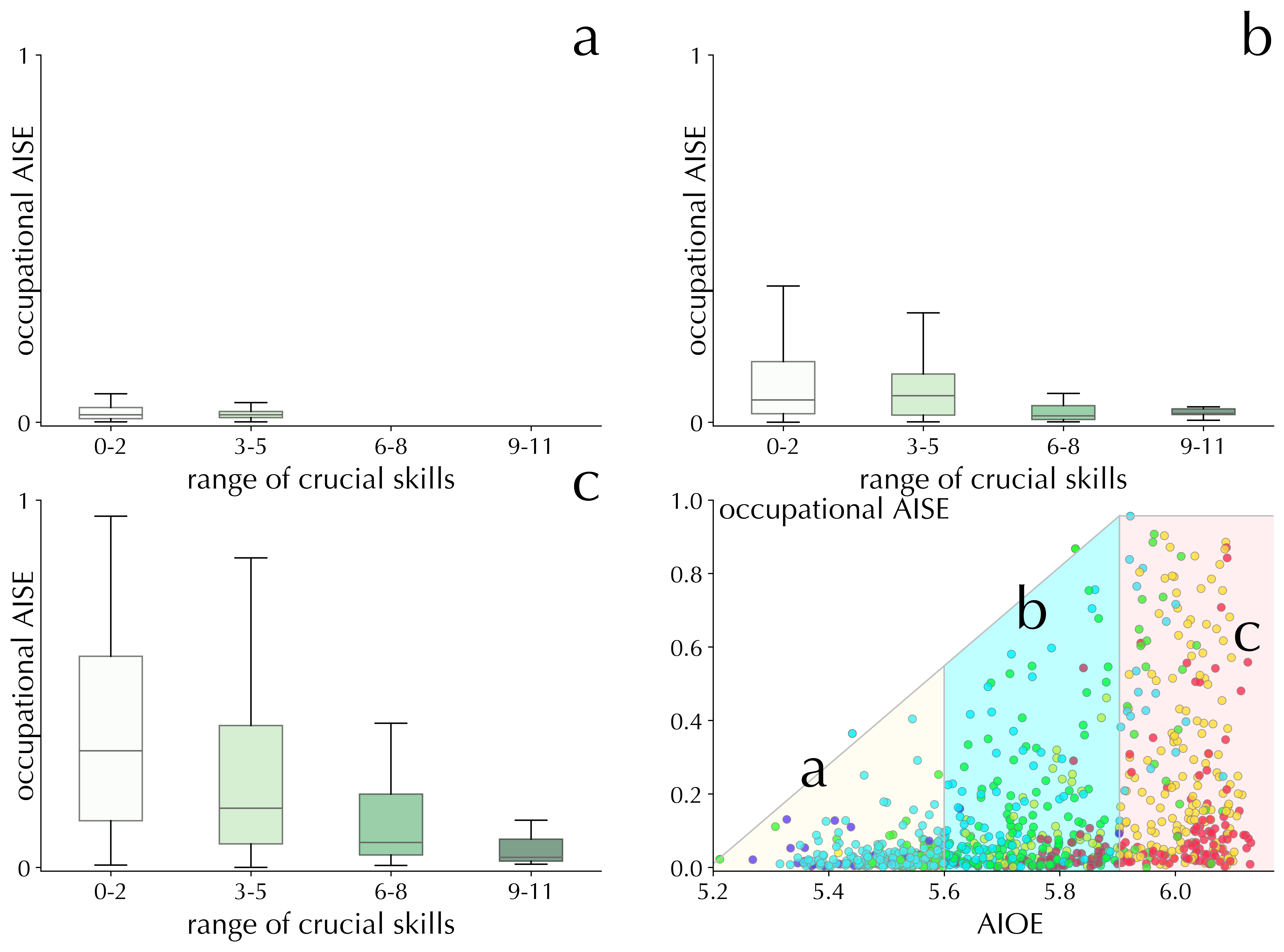}
    \caption{\textbf{\startupexposure vs number of skills with importance larger than 4 for different fixed ranges of AIOE values.} The width of the violin is proportional to density of jobs within the corresponding interval of Y Combinator values and the black dot represents the average. The number of points for each violin is denoted by $n$.}
    \label{fig:5}
\end{figure}

\clearpage
\sisubsection{Skill importance in the \startupexposure-AIOE space}
In the main text, we analyzed the frequency of required skills with an O*NET importance score exceeding 4 for jobs in regions characterized by high AIOE and high \startupexposure, as well as high AIOE and low \startupexposure. The following figure displays a similar bar plot, but with the importance threshold for required skills lowered to 3. The findings are consistent with those presented in the main text. Jobs in the lower-right region of the AIOE-\startupexposure space show a significantly higher likelihood of requiring skills with an importance score above 3. Additionally, skills related to social and human domains, such as \textit{social perceptiveness} and \textit{instructing}, are more frequently observed in this area.

The next two figures, instead, display the frequency of required skills with an O*NET importance score exceeding 4 or 3 for jobs in regions characterized by high AIOE and low \startupexposure, as well as low AIOE and low \startupexposure. Although tasks for jobs in these regions have not been prominently targeted by AI-based startups, they show significant differences in potential AI exposure. Both bar plots clearly demonstrate that jobs in the lower-left corner of the \startupexposure-AIOE space are less complex, at least regarding the skill sets recorded in O*NET, compared to jobs with higher AIOE. This is evident from the fact that only few of these jobs require skills with high importance values.

\begin{figure}[htbp]
    \centering
    \includegraphics[width=1\textwidth]{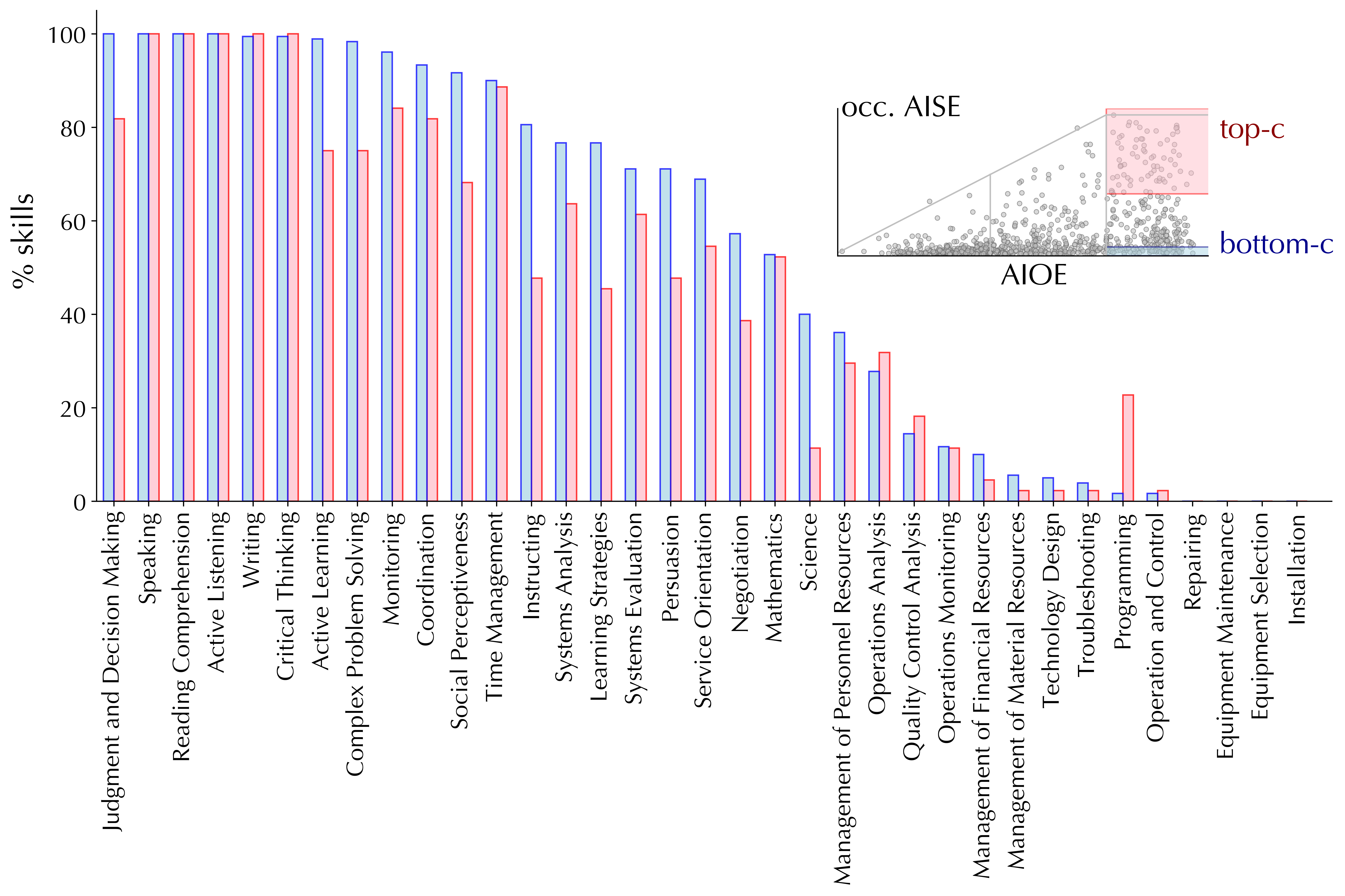}
    \caption{\textbf{Frequency of skills with importance larger than 3 for two different region of the \startupexposure-AIOE space.} Blue bars describe the jobs in the top right part of the YCombinator-AIOE space; red bars describe the jobs in the bottom right part of the YCombinator-AIOE space. For each skill, the smaller bar is over the higher bar..}
    \label{fig:10}
\end{figure}

\begin{figure}[htbp]
    \centering
    \includegraphics[width=1\textwidth]{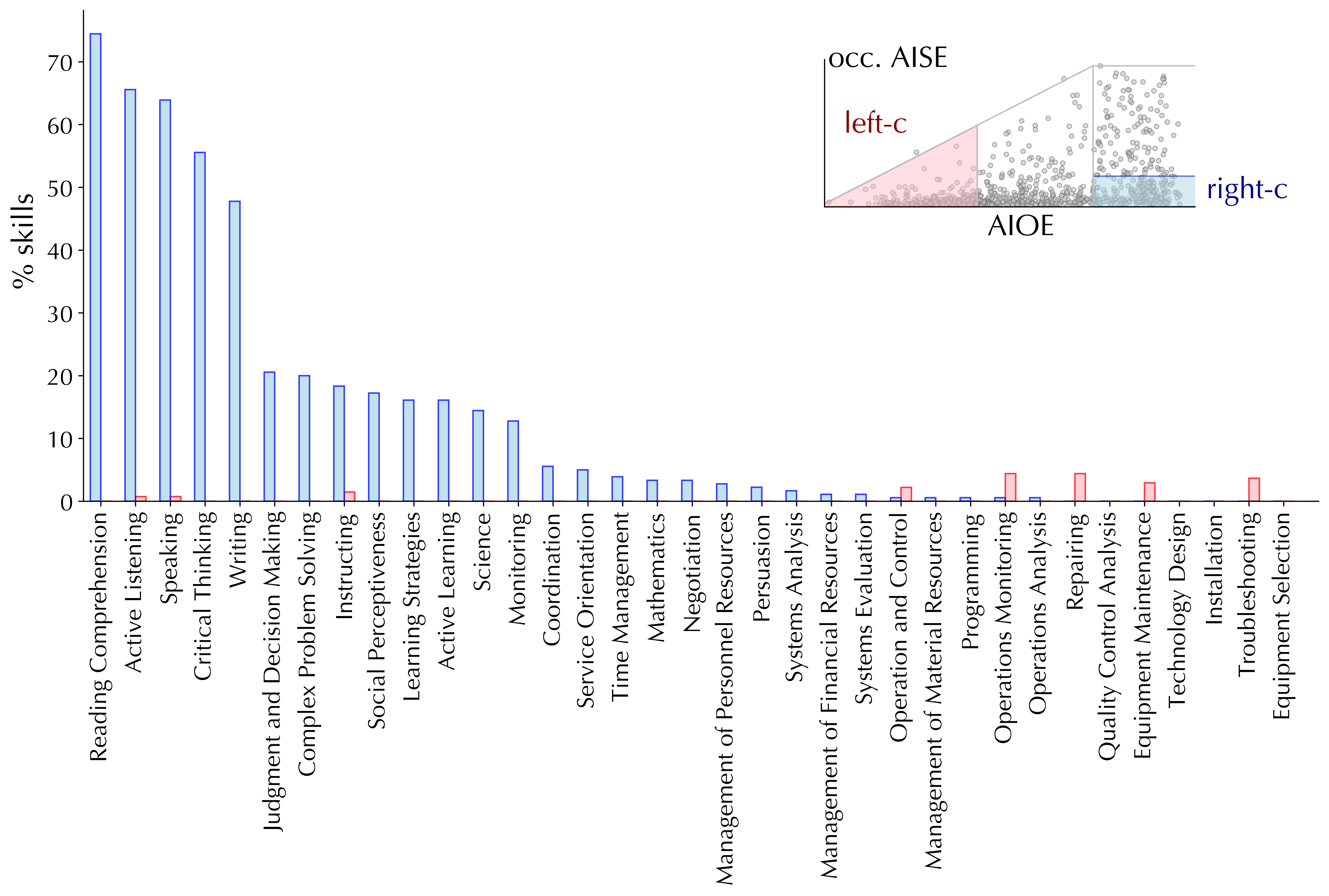}
    \caption{\textbf{Frequency of skills with importance larger than 4 for two different region of the \startupexposure-AIOE space.} Blue bars describe the jobs in the bottom left part of the YCombinator-AIOE space; red bars describe the jobs in the bottom right part of the YCombinator-AIOE space. For each skill, the smaller bar is over the higher bar.}
    \label{fig:11}
\end{figure}

\begin{figure}[htbp]
    \centering
    \includegraphics[width=1\textwidth]{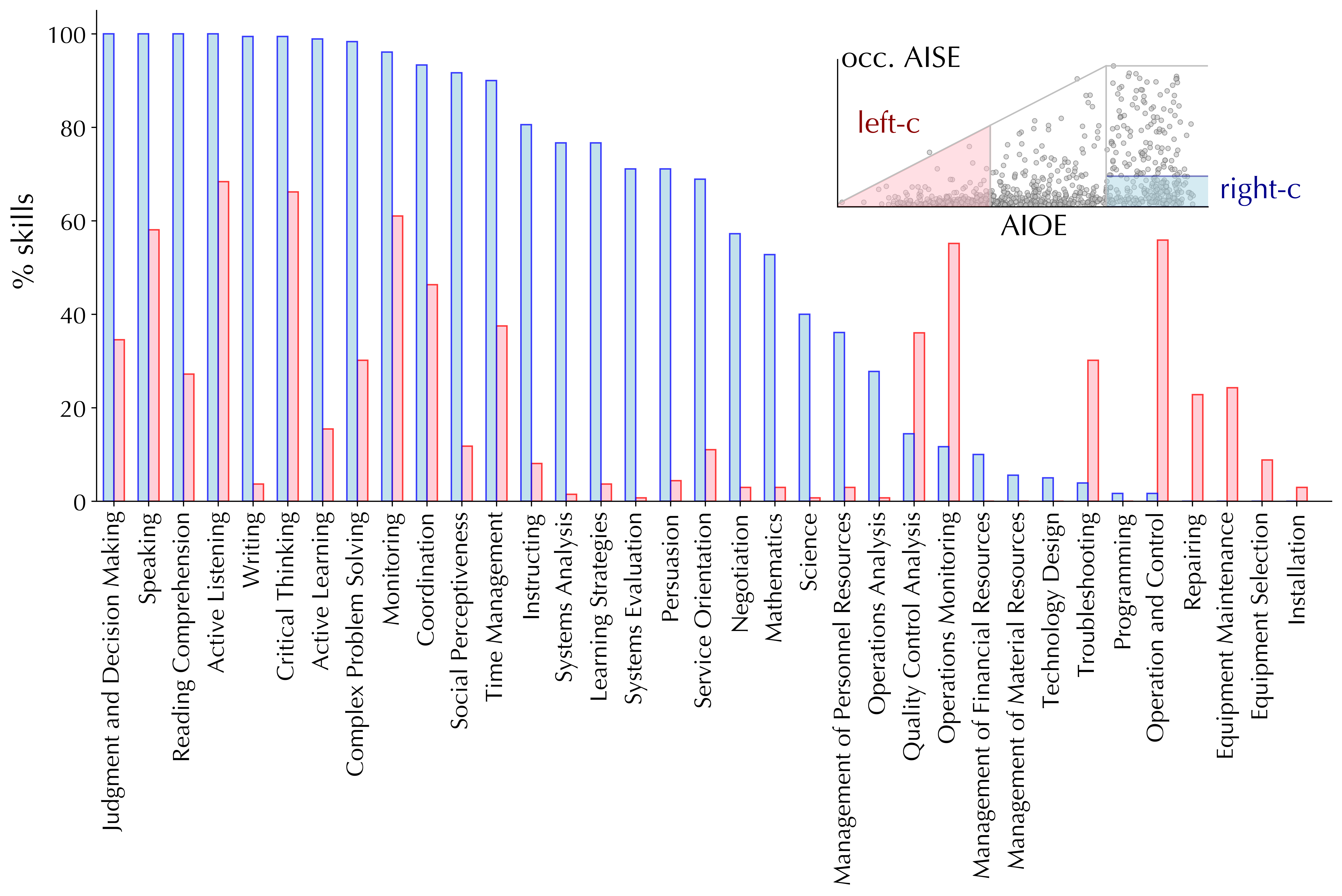}
    \caption{\textbf{Frequency of skills with importance larger than 3 for two different region of the \startupexposure-AIOE space.} Blue bars describe the jobs in the bottom left part of the YCombinator-AIOE space; red bars describe the jobs in the bottom right part of the YCombinator-AIOE space. For each skill, the smaller bar is over the higher bar.}
    \label{fig:12}
\end{figure}

\clearpage
\sisubsection{Test with different input and prompts}
As described in the methods section, to evaluate the robustness of our methodology and findings, we repeated the experiment under two different conditions. In the first scenario, we feed Llama3 with the same prompt as before but replaced the detailed descriptions of the startups with shorter ones provided by Y Combinator.

For example, while the detailed description for the startup \textit{Studdy} is :\\
"At Studdy, our mission is to unlock the full potential of the next generation by providing a personalized AI tutor for every student.
Studdy Buddy is a multilingual tutor that uses AI, as well as advanced speech, text, and image recognition technology to supercharge students' ability to learn new subjects.
We believe that making self-learning as easy as possible for as many students as possible (no matter their cultural, social, or educational background) is the key to unlocking the full potential of students around the world.
We're a passionate team of AI experts, educators, and builders - if you also have a passion for transforming education we'd love to hear from you.
Shoot us a message at team@studdy.ai!".\\

Instead, the short description is: "An AI math tutor for every student".

\begin{figure}[thbp]
	\centering
	\includegraphics[width=0.7\textwidth]{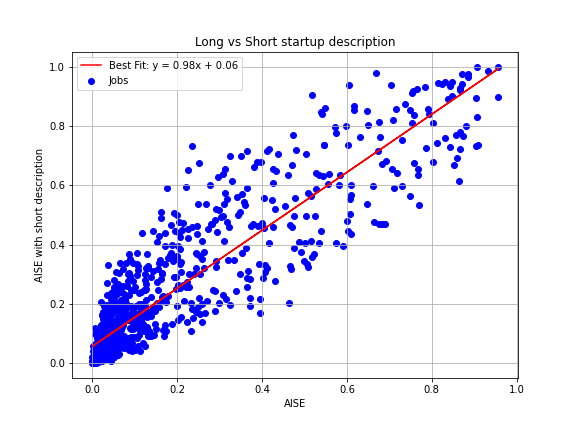}
	\caption{\textbf{Scatter plot comparing the standar occupational AISE with the AISE constructed with the short startup description provided by Y Combinator}.}
	\label{fig:short}
\end{figure}

While the detailed descriptions are more informative, they often include extraneous details about the founders or funding. On the other hand, the shorter descriptions are more focused on the product being developed, minimizing the likelihood of Llama3 being influenced by irrelevant details. Nevertheless, as shown in Fig. \ref{fig:short}, the results with the shorter descriptions closely align with those obtained from the detailed descriptions. In particular, the pearson correlation coefficient is 0.91, while the Kendall coefficient, a metric that measures the ordinal association between two variables, is 0.74.

\begin{figure}[thbp]
	\centering
	\includegraphics[width=0.7\textwidth]{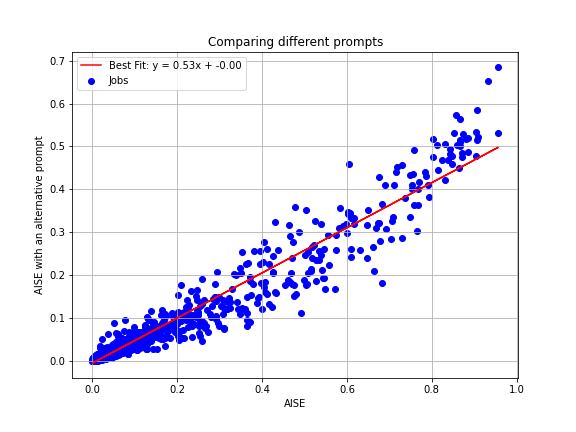}
	\caption{\textbf{Scatter plot comparing the standar occupational AISE with the AISE constructed by feeding Llama3 with a different prompt}.}
	\label{fig:q2}
\end{figure}

In the second scenario, we revised the prompt as follows:\\
\\
$\{$\textit{“role”: “system”, “content”: “You are an AI specialist.”}$\}$ \\
$\{$\textit{“role”: “user”, “content”: “Given the following startup description: ” + startup[j] + “and given the following job description: “+ job[i] + “Is the product or service developed by the startup designed to directly replace humans to perform some of the described job’s tasks? Use only the information provided in the two descriptions. Reply only with yes or no.”}$\}$\\
\\
This revised prompt takes a more direct approach by explicitly asking whether the startup's product or service is designed to replace human L’abiura for specific tasks mentioned in the job description. In contrast, the previous prompt gave Llama 3 more interpretive freedom to assess the relationship between the startup's AI product/service and the job's tasks. Again, despite these differences in framing, the exposure rankings generated by the two prompts remain highly consistent, as shown in Fig.\ref{fig:q2}. In particular the Pearson correlation coefficient is 0.97, while the Kendall coefficient is 0.85.

\clearpage

\begin{figure}[thbp]
	\centering
	\includegraphics[width=1\textwidth]{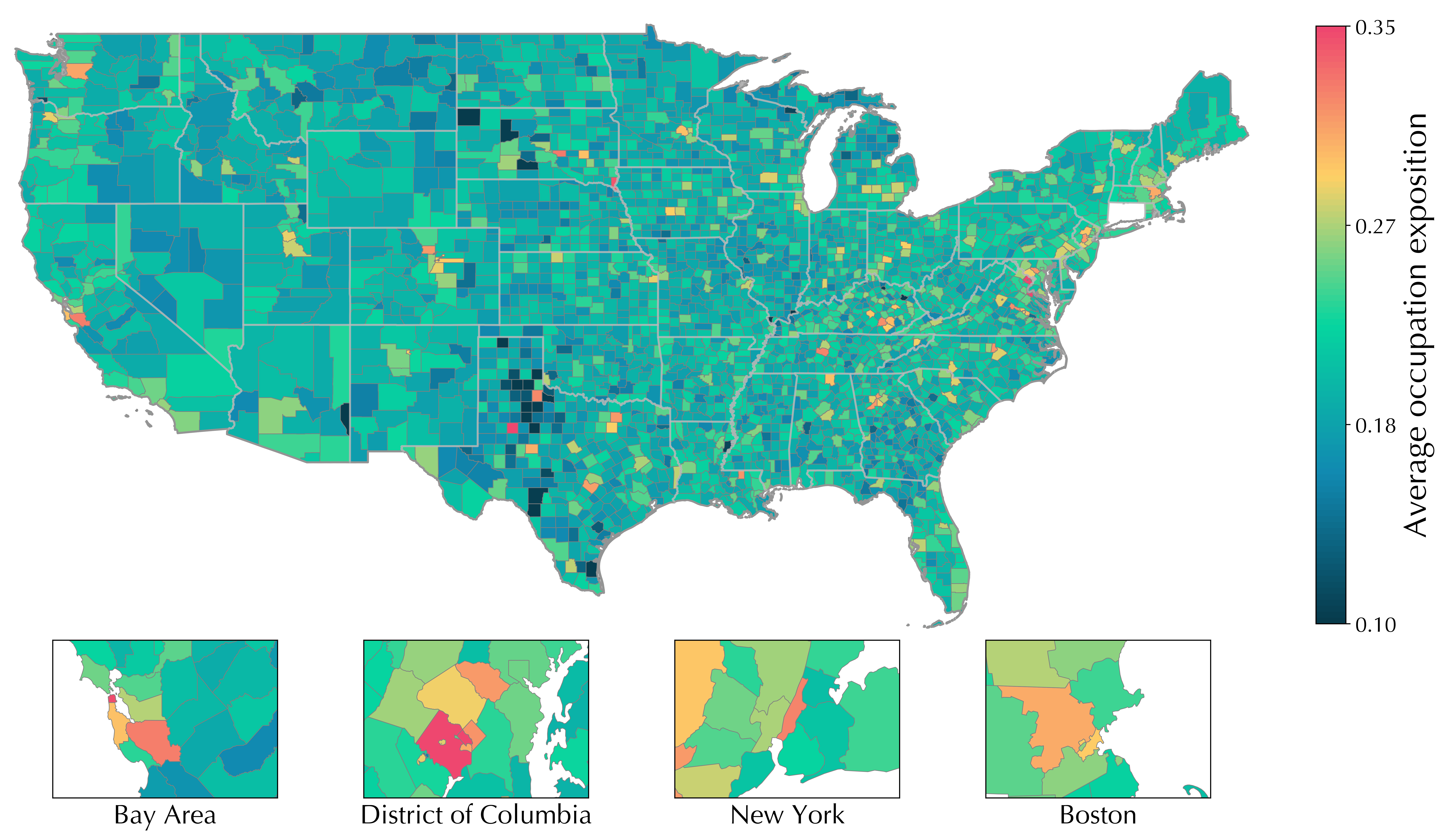}
	\caption{\textbf{Map of average exposure to AI of the workforce of the American counties}.
 The color-code of the figure indicates the Average Occupation Exposition of the counties according to the \startupexposure. At the bottom panels, the four most exposed counties are zoomed.
	}
	\label{fig:us_map_1}
\end{figure}

\sisubsection{US County map}
For completeness, we show in Figure \ref{fig:us_map_1} the average Occupational \startupexposure of counties. 
The occupational data at the county level is not available, and it is obtained by projecting the data from the national level on the sectorial distribution of the countries.
This approach is the one followed by~\cite{felten2018method} and covers the whole American surface, at the price of more noisy and less precise information.
Countries that have a low population tend to have more uncertainty on the exposition measure, such as in the central Texas.
However, the information provided is consistent with that observed in the map in Figure \ref{fig:us_map} of the main text.

%%% Ditemi quale vi piace di più...

\clearpage
\sisubsection{Robotic Startup Exposure}
In this section, we demonstrate how the methodology used to compute the AISE can be generalized to study the forward-looking impact of robotics and its integration with AI on the labour market. While there is a long-standing body of literature on the impact of robotics and automation \citep{acemoglu2011skills, acemoglu2020robots, graetz2018robots}, interest in the impact of AI integration with robotics is more recent, with many believing that it will have an even more radical impact than AI alone \citep{soori2023artificial, barbieri2020testing}. As we show below, our framework allows us to observe directly the impact of the joint combination of AI and robotic, as most of the Y Combinator startups with a robotic-related tag are integrating AI software with robot's hardware.

Adopting the same strategy we used to build AISE, we feed Llama3 the SOC occupations' textual descriptions and the descriptions of all the AI startups  with a robotics-related tag on Y Combinator -- see Section \ref{sec:data&methods} for more details on the technical procedure. Therefore, for each occupation, we define the Startup Robotic Exposure (RSE) index as the normalized number of startups developing robotic applications identified by the LLM as substitutes for one or more of the essential tasks described in the O*NET short descriptions. It is important to notice the exploratory nature of this analaysis, in fact only about 100 startups with a robotics-related tag are present in the dataset, compared to approximately 1000 AI-tagged startups. 

\begin{figure}[thbp]
	\centering
	\includegraphics[width=1\textwidth]{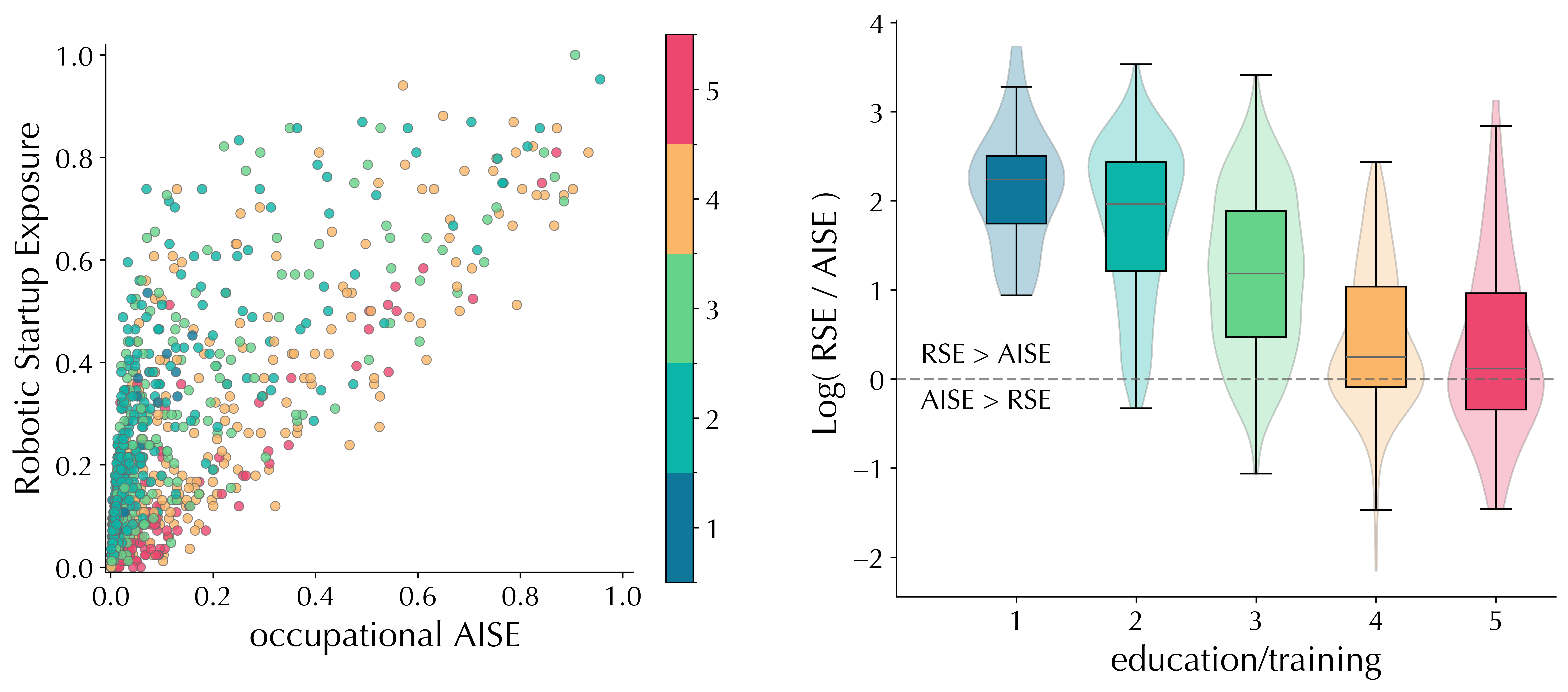}
	\caption{\textbf{Comparison of the Robotic Startup Exposure with AISE and the educational levels}.
    The left panel shows the scatter plot of the Robotic Startup Exposure with the occupational AISE. Each dot represents an O*NET job and the color code indicates the educational/training level of the job accessible from the O*NET database.
    The right panel shows the barplot of the relative importance of the two exposure measures subdivided by education/training level.
    }
	\label{fig:RSE}
\end{figure}

Figure \ref{fig:RSE} details our findings. The left panel for each occupation shows a AISE versus RSE scatter plot, colour-coded according to job zones. Interestingly and unsurprisingly, when robotic is considered, the pattern of exposure change. In fact, several occupations with low AISE scores actually display a high RSE. As illustrated in the right panel, this is particularly evident for professions in lower job zones, which require more manual abilities or skills (e.g., \textit{Control Movement Abilities}). Instead, professions in job zone 5, thus requiring high levels of educations, present low values for both AISE and RSE.

A peculiar finding is that jobs with high AISE also have high RSE, despite these jobs (such as \textit{Office Clerk, General}) do not require any physical skills. This is because, as already mentioned, in the Y Combinator dataset, most startups with a robotics-related tag also have an AI-related tag\footnote{This also explains why there are only about a hundred startups with robotics-related tags. Indeed, the technology that combines AI and robotics is still in its early stages, and there is still significant uncertainty regarding its reliability and adoption \citep{eloundou2024gpts}.}. Therefore, we are observing startups developing AI products integrated into hardware, and Llama 3 considers a job exposed to these startups even if it is exposed only to the AI software component of the products they develop.

Overall, these preliminary results suggest that the joint action of AI and robotics will be pervasive across all occupations, warranting further in-depth studies.

\clearpage

\end{document}